\documentclass[letterpaper,showpacs,twocolumn,superscriptadress,floatfix]{revtex4-1}

\usepackage[utf8]{inputenc} 
\usepackage{latexsym} 
\usepackage{amsmath}  
\usepackage{amssymb}
\usepackage{hyperref}
\usepackage{color,graphicx}
\usepackage{epsfig}
\usepackage{epstopdf}

\begin{document}

\newcommand{\vphi}{\varphi}
\newcommand{\vphic}{\varphi^\ast}
\newcommand{\Pic}{\tilde{\Pi}}
\newcommand{\Picc}{\tilde{\Pi}^\ast}
\newcommand{\chic}{\tilde{\chi}}
\newcommand{\chicc}{\tilde{\chi}^{\ast}}
\newcommand{\chicci}{\tilde{\chi}^{\ast i}}

\newcommand{\h}{\mathfrak{h}}

\newcommand{\hg}{\hat{\gamma}}
\newcommand{\hG}{\hat{\Gamma}}
\newcommand{\hA}{\hat{A}}
\newcommand{\hR}{\hat{R}}
\newcommand{\hD}{\hat{\Delta}}
\newcommand{\hDcov}{\hat{\D}}
\newcommand{\hnabla}{\hat{D}}

\newcommand{\fg}{\mathring{\gamma}}
\newcommand{\fG}{\mathring{\Gamma}}
\newcommand{\fR}{\mathring{R}}
\newcommand{\fnabla}{\mathring{D}}

\newcommand{\p}{\partial}
\newcommand{\bad}{\beta^m \p_m}
\newcommand{\divB}{\hnabla_m \beta^m}
\newcommand{\sdivB}{\sigma\hnabla_m \beta^m}
\newcommand{\ssdivB}[1]{\frac{2}{3}\sigma {#1}\,\hnabla_m \beta^m}
\newcommand{\D}{\mathcal{D}}

\newcommand{\Lie}[2]{\pounds_{\vec{#1}}{#2}}

\newcommand{\miguel}[1]{\textcolor{red}{{\bf #1}}}
\newcommand{\jm}[1]{\textcolor{blue}{{\bf #1}}}

\title{Gravitational collapse of charged scalar fields}

\author{Jose M. Torres}
\email{jose.torres@nucleares.unam.mx}

\author{Miguel Alcubierre}
\email{malcubi@nucleares.unam.mx}

\affiliation{Instituto de Ciencias Nucleares, Universidad Nacional
Aut\'onoma de M\'exico, A.P. 70-543, M\'exico D.F. 04510, M\'exico.}

\date{\today}

\begin{abstract}
In order to study the gravitational collapse of charged matter we
analyze the simple model of an self-gravitating massless scalar field
coupled to the electromagnetic field in spherical symmetry. The
evolution equations for the Maxwell-Klein-Gordon sector are derived in
the 3+1 formalism, and coupled to gravity by means of the
stress-energy tensor of these fields. To solve consistently the full
system we employ a generalized Baumgarte-Shapiro-Shibata-Nakamura
(BSSN) formulation of General Relativity that is adapted to spherical
symmetry. We consider two sets of initial data that represent a time
symmetric spherical thick shell of charged scalar field, and differ by
the fact that one set has zero global electrical charge while the
other has non-zero global charge. For compact enough initial shells we
find that the configuration doesn't disperse and approaches a final
state corresponding to a sub-extremal Reissner-N\"ordstrom black hole
with $|Q|<M$. By increasing the fundamental charge of the scalar field
$q$ we find that the final black hole tends to become more and more
neutral. Our results support the cosmic censorship conjecture for the
case of charged matter.
\end{abstract}


\maketitle

\section{Introduction}
\label{sec:intro}

Gravitational collapse is a generic feature of nature and occurs
whenever gravitation dominates over the repulsive interactions that
take place inside matter. In the context of General Relativity this
phenomena is very important since it is the mechanism that drives the
formation of singularities in the spacetime itself, and understanding
the causal structure of such singularities is of prime importance to
ensure the predictability of the theory. It is widely believed that
the final fate of gravitational collapse is the formation of a black
hole in which the singularities are casually disconnected from distant
observers. The fact that no spacetime singularities have been
identified has led to establish the cosmic censorship
conjecture~\cite{Penrose99}, which states that no naked singularities
can develop from generic initial data where the spacetime is regular
and the matter satisfies the dominant energy condition.  Beyond
General Relativity, it is generally expected that a description that
accounts for the quantum nature of gravity and/or spacetime should
avoid the formation of singularities. Even in that case it is
important to understand the dynamics of gravitational collapse since
quantum effects should only play a role in those regions where a
singularity is expected to form classically, so that for macroscopic
black holes the effective dynamics near the horizon should not depart
significantly from those predicted by the classical theory.

It is known that the only stationary, asymptotically flat solutions of
the Einstein equations that represent a black hole correspond to the
Kerr-Newman family which is characterized by the mass, electrical
charge and angular momentum. These solutions only represent black
holes when the parameters satisfy the relation $M^2>Q^2+a^2$ (with $M$
the mass, $Q$ the total electric charge and $a$ the angular momentum
per unit mass), otherwise there is no event horizon present in the
spacetime and the singularity is naked. In a realistic scenario of
gravitational collapse it is expected that the outer region of the
spacetime will correspond asymptotically to this solution, but in
principle there is no restriction that ensures the above relation
would hold. There have been numerical and analytical studies of
collapsing scenarios that lead to naked singularities
\cite{Shapiro91,Choptuik93,Christodoulou94}, but it has not been
asserted that these examples pose a real threat to the cosmic
censorship conjecture since these results may come from the symmetries
imposed and even from the gauge conditions. It seems that, at least
for the uncharged case $q=0$, the generic outcome of gravitational
collapse is a black hole and the exceeding angular momentum contained
in the initial data is lost due to gravitational radiation.

It is also interesting to consider the collapse of charged matter
which could lead to the formation of charged black holes or even naked
singularities. It is expected that the collapse of charged matter
would not lead to a super-extremal solution ($Q \ge M$) since in this
case the electric repulsion is comparable to the gravitational
attraction. A simple model for charged matter is that of a complex
scalar field coupled to the electromagnetic field, this configuration
represents matter made of charged bosonic particles and its
antiparticles which have opposite charge. Scalar field matter has been
widely used in General Relativity because its evolution is governed by
a simple wave equation and does not develop discontinuities as it
happens with fluids. Self-Gravitating complex scalar fields may form
boson stars, which are regular self-gravitating configurations of
scalar field that do not disperse and have properties that resemble
those of neutron stars (a recent review can be found
in~\cite{Liebling:2012fv}). Boson stars can be made also out of
charged scalar field~\cite{Jetzer89}, providing an example of an
astrophysical system that posses net charge and is susceptible to
collapse.  The numerical simulation of the collapse of charged scalar
fields with spherical symmetry has been previously studied with many
different goals: For example, in~\cite{Petryk05} a standard $3+1$
decomposition is used with polar slicing to analyze critical
phenomena, critical behavior similar to the one occurring in the
non-charged case~\cite{Choptuik93} is found and the authors show that
the charge approaches zero faster than the mass as one approaches the
critical solution, resulting in a critical solution that has no
charge; in~\cite{Oren03} a null formalism is introduced to probe the
internal structure of the resulting black hole. In this paper we are
interested on the resulting configuration as viewed by outside
observers, in particular we want to determine if the outcome of this
process can be a naked singularity.

This paper is organized as follows: in Section~\ref{sec:problem} we
recast the evolution equations of the matter in spherical symmetry and
show the terms of the stress-energy tensor that couple to the
evolution equations for the geometric variables. In
Section~\ref{sec:idata} we construct initial data that represents
collapsing charged matter. Then, in Section~\ref{sec:numerical} we
discuss some analysis tools and some features of the numerical
implementation. In Section~\ref{sec:results} we present the results of
our simulations. We conclude in Section~\ref{sec:conclusions}.

\section{Field Equations}
\label{sec:problem}

\subsection{Einstein-Maxwell-Klein-Gordon system in spherical symmetry}
\label{sec:EMKG}

The study of the collapse of a charged scalar field is addressed by
integrating the full Einstein-Maxwell-Klein-Gordon system of evolution
equations. We will use the 3+1
formalism~\cite{Arnowitt62,York79,Alcubierre08a}, and in particular
adopt the generalized
Baumgarte-Shapiro-Shibata-Nakamura~\cite{Shibata95,Baumgarte:1998te}
(BSSN) formulation in curvilinear
coordinates~\cite{Brown:2009dd,Alcubierre:2010is} to recast the
Einstein equations as an initial value problem in spherical symmetry.
The spacetime metric is then written as
\begin{equation}
ds^2 = \left( - \alpha^2 + \beta_i \beta^i \right) dt^2
+ 2 \beta_i \, dt dx^i + \gamma_{ij} \: dx^i dx^j \: ,
\end{equation}
with $\alpha$ the lapse function, $\beta^i$ the shift vector, and
$\gamma_{ij}$ the spatial metric (notice that indexes of spatial
quantities are lowered and raised with the spatial metric
$\gamma_{ij}$).

The matter, including both the electromagnetic and scalar
fields, couples to this system only through contributions from the
stress-energy tensor. We require only to analyze the evolution
equations of the matter fields in the curved spacetime characterized
by the spatial metric $\gamma_{ij}$ and extrinsic curvature $K_{ij}$

In spherical symmetry the spatial line element can be written without
loss of generality as
\begin{equation}
dl^2 = \psi(r,t)^4 \left( A(r,t) dr^2 + r^2 B(r,t) d\Omega^2 \right) \; .
\label{eq:metSpheric}
\end{equation}
with $\psi$ a conformal factor and $d\Omega^2$ the standard solid
angle element.

\subsection{Electromagnetic field}
\label{sec:EM_evol}

The evolution equations for the electromagnetic field are naturally
formulated in terms of the electric and magnetic fields measured by
the Eulerian observers~\cite{Alcubierre:2009ij}
\begin{eqnarray}
\label{eq:Electric}
E^\mu &:=& -n_\nu F^{\nu\mu} \; ,\\
\label{eq:Magnetic}
B^\mu &:=& -n_\nu F^{*\nu\mu} \; ,
\end{eqnarray}
with $n^\mu$ the unit normal future-pointing vector to the spatial
hypersurfaces.

When projecting the covariant Maxwell equations onto the normal vector
to the spatial hypersurfaces one finds two constraint equations
\begin{eqnarray}
\label{eq:Max3}
D_i E^i &=& 4\pi \rho_e \; , \\
\label{eq:Max4}
D_i B^i &=& 0 \; ,
\end{eqnarray}
with $D_i$ the covariant derivative compatible with the spatial metric
$\gamma_{ij}$, and where $\rho_e=-n_\mu j^\mu$ is the charge density
measured by the Eulerian observers.

On the other hand, when projecting onto the hypersurfaces we obtain
the evolution equations for the electric and magnetic fields
\begin{eqnarray}
\label{evE}
\frac{d}{dt} \: E^i &=& (D \times \alpha B)^i - \alpha K E^i
- 4 \pi \alpha \: {}^{(3)}\!j_{\rm e}^i \; , \\
\label{evB}
\frac{d}{dt} \: B^i &=& - (D \times \alpha E)^i + \alpha K B^i \; ,
\end{eqnarray}
where \mbox{${}^{(3)}\!j_{\rm e}^i := \gamma_{\,\,\,\mu}^{i} j^\mu$}
is the current density measured by the Eulerian observers, $d/dt=\p_t
- \Lie{\beta}{}$ is the derivative in the normal direction to the
spatial hypersurfaces, and where the rotational operator acting on a
vector $v^i$ is defined as $\left( D \times \, v \right)^i :=
\epsilon^{imn} D_m \,v_n$, with $\epsilon^{imn}:=n_\mu \epsilon^{\mu
  imn}$ the Levi-Civita tensor on the spatial hypersurfaces.

Since the charged scalar field couples directly to the electromagnetic
potential $A^\mu$, we also need to specify the evolution equation for
the potential. To do this we use the following $3+1$ decomposition of
this vector field
\begin{eqnarray}
\label{eq:potEM}
\Phi &:=& -n_\mu A^\mu \; , \\
a^i &:=& {}^{(3)}\!A^i = \gamma^{i}_\mu A^\mu \; ,  
\end{eqnarray}
which defines the scalar and vector electromagnetic potentials
respectively as measured by Eulerian observers. Writing the Faraday
tensor in terms of the electromagnetic potentials as
\begin{equation}
F_{\mu\nu}:=\p_\mu A_\nu - \p_\nu A_\mu \; ,
\end{equation}
the definition of the electric field~\eqref{eq:Electric} becomes an
evolution equation for the vector potential
\begin{equation}
\label{eq:potvec}
\frac{d}{dt} \: a_i = - \alpha E_i - D_i (\alpha \Phi) \; ,
\end{equation}
while the definition of the magnetic field~\eqref{eq:Magnetic} becomes
\begin{eqnarray}
B^i &=& \frac{1}{2} \: \epsilon^{imn}
\left( \partial_m a_n - \partial_n a_m \right) \nonumber \\
&=& \epsilon^{imn} \partial_m a_n
= \left( D \times a \right)^i \: .
\label{eq:Brota}
\end{eqnarray}

In order to find the scalar potential at all times it is now necessary
to fix the gauge freedom of the theory. Here and in what follows we
choose the Lorentz gauge $\nabla_\mu A^\mu = 0$, which when written in
$3+1$ language becomes an evolution equation for the scalar potential
\begin{equation}
\label{eq:potsca}
\frac{d}{dt} \: \Phi = \alpha K \Phi - D_i \left( \alpha\, a^{i} \right) \; .
\end{equation}

When one considers the case of spherical symmetry the above equations
simplify considerably since all vector fields can have only a radial
component. This then implies, along with equation~\eqref{eq:Brota},
that the magnetic field vanishes, so in practice we only have to
consider the equations for the scalar potential and the radial
component of both the vector potential and the electric field, which
take the form
\begin{eqnarray}
\partial_t \Phi &=& \beta^r \partial_r \Phi + \alpha K \Phi \nonumber \\
&-& \frac{1}{r^2 \psi^6 \sqrt{A}B} \; \partial_r
\left( \frac{ \alpha B r^2 \psi^2}{\sqrt{A}} \: a_r \right) \; ,
\label{eq:spotsca} \\
\partial_t a_r &=& \beta^r \partial_r  a_{r} 
+ a_r \partial_r \beta^r - \alpha A \psi^4 E^r
- \partial_r \left( \alpha \Phi \right) \; , \hspace{8mm}  
\label{eq:spotvec} \\
\partial_t E^r &=& \beta^r \partial_r  E^r 
- E^r \partial_r \beta^r + \alpha K E^r - 4 \pi j_{\,em}^r \; ,
\label{eq:selec}
\end{eqnarray}
with $\beta^r$ the radial component of the shift vector.

\subsection{Charged scalar field}
\label{sec:chsf_evol}

For the charged scalar field we can do a similar 3+1 decomposition.
We start from the Lagrangian
\begin{equation}
L = -\frac{1}{2}\left[ (\D_\mu \varphi)^\ast \D^\mu \varphi
+ 2 V(|\varphi|^2) \right] \; ,
\label{eq:lagC}
\end{equation}
where $\D_\mu= \nabla_\mu + iq A_\mu $ is the gauge invariant
derivative, with $\nabla_\mu$ is the covariant derivative adapted to
the spacetime metric, and where $V(|\varphi|^2)$ is a self-interaction
potential for the scalar field (in all our numerical simulations below
we always take $V=0$, but we consider it here for generality). We say
that $\D_\mu$ is gauge invariant in the sense that its application on
the scalar field is not altered under a transformation of the form
\begin{eqnarray}
\vphi &\rightarrow& e^{-iq\vartheta(x^\alpha)} \vphi \; , \\
A_\mu &\rightarrow& A_\mu + \p_\mu \vartheta(x^\alpha) \; ,
\end{eqnarray}
with $\vartheta(x^\alpha)$ an arbitrary scalar function of spacetime.

The complex scalar field has two degrees of freedom, so we can treat
it equivalently as two independent variables corresponding to the real
and imaginary parts, or as a complex variable and its complex
conjugate. Variation of the action while fixing $\vphi$ gives the
following evolution equation
\begin{equation}
\label{eq:KGC1}
\left(\D^\mu \D_\mu - 2 V^\prime \right) \varphi = 0 \; , \quad 
V^\prime := \frac{d}{d(|\vphi|^2)} \: V(|\vphi|^2) \; .
\end{equation}
From the gauge invariance of the scalar field it follows that there is
a conserved current which acts as the source of the electromagnetic
field
\begin{equation}
\label{eq:CurrentC}
(j_{\vphi})^\mu = \frac{i}{2} \left( \varphi^\ast \D^\mu \varphi
- (\D^\mu \varphi)^\ast\varphi \right) \; .
\end{equation}
This implies that one has a conserved quantity when integrating the
projected component over a spacelike hypersurface. This conserved
quantity is the total ``boson number'' $N$ and is proportional to the
total electric charge $Q$ as $Q=qN$, with $q$ the electric charge of
the fundamental particles. Thus, for consistency, the electromagnetic
current has the form
\begin{equation}
\label{eq:CurrentEM}
(j_{{\rm e}})^\mu = \frac{iq}{2} \left( \varphi^\ast \D^\mu \varphi -
(\D^\mu \varphi)^\ast\varphi \right) \; ,
\end{equation}

The evolution equation~\eqref{eq:KGC1} is in fact a wave equation with
nonlinear sources on the electromagnetic potentials.  Expanding the
gauge invariant derivatives we obtain
\begin{eqnarray}
\label{eq:KGC2}
\Box \vphi &=& 2 V^\prime \vphi - q \left( 2 i A^\mu \nabla_\mu \vphi
\right. \nonumber \\
&+& \left. i \vphi \nabla^{\mu} A_\mu - q A_\mu A^\mu \vphi \right) \; .
\end{eqnarray}
Although in the Lorentz gauge one of the source terms above vanishes,
one can see that the two degrees of freedom are coupled by the
interaction term $2 iq A^\mu \nabla_\mu \vphi$, while the quadratic
term on the electromagnetic potential gives rise to an effective mass
for the scalar field. The equation above can be reduced to a first
order system by defining the variables
\begin{eqnarray}
\Pi &:=& n^\mu \nabla_\mu \varphi
= \frac{1}{\alpha} \left( \partial_t \varphi - \beta^i \partial_i \varphi \right)
\; , \label{eq:Pi} \\
\chi_i &:=& \nabla_i \varphi = \partial_i \varphi \; . \label{eq:Chi}
\end{eqnarray}

In terms of these variables the evolution equation above turns out to
be equivalent to the system
\begin{eqnarray}
\label{eq:ev1}
\frac{d}{dt} \: \varphi &=& \alpha \Pi \; , \\
\label{eq:ev2}
\frac{d}{dt} \: \chi_i &=& D_i(\alpha \Pi) + \beta^j\left( D_i \chi_j
- D_j \chi_i \right) \; , \\
\label{eq:ev3}
\frac{d}{dt} \: \Pi &=&  D_i (\alpha \chi^i) + \alpha \Pi K \nonumber \\
&-& \alpha \left[ 2 V^\prime + q^2 \left( a_i a^i - \Phi^2 \right)
- i q \nabla^{\mu} A_\mu \right] \vphi \nonumber \\ 
&+& 2i q \alpha \left( a^i \chi_i + \Phi \Pi \right) \; .
\end{eqnarray}

When considering the case of spherical symmetry we are left only with
the radial derivative $\chi \equiv \chi_r$, and the above equations
reduce to
\begin{eqnarray}
\label{eq:sphi}
\partial_t \varphi &=& \alpha \Pi + \beta \chi \; , \\
\label{eq:schi}
\partial_t \chi &=& \partial_r \left( \alpha \Pi + \beta\chi \right) \; , \\
\label{eq:spi}
\partial_t \Pi &=& \beta \partial_r \Pi
+ \frac{1}{r^2 \psi^6 \sqrt{A}  B }\partial_r 
\left( \alpha\frac{B r^2 \psi^2}{\sqrt{A}} \chi \right) 
\nonumber \\
&+& \alpha \Pi K - \alpha \left[ 2 V^\prime 
+ q^2 \left( \frac{a_{r}^2}{A \psi^4 }
- \Phi^2 \right) \right] \vphi \nonumber \\
&+& 2i q \alpha \left[   
\frac{a_r \chi}{A \psi^4} + \Phi \Pi  \right] .
\end{eqnarray}

It turns out that it is also useful to define the gauge invariant
versions of the variables $\Pi$ and $\chi_i$ since the source terms
appearing in the equations for the gravitational and electromagnetic
fields take simpler forms in terms of them:
\begin{eqnarray}
\label{eq:auxC}
\Pic &:=& n^\mu \D_\mu \vphic = \Pi -iq\Phi\vphi \; , \\
\chic_i &:=& \gamma^\mu_i \D_\mu \vphi = \chi_i + i q a_i \vphi .
\end{eqnarray}

In terms of these variables we can now define the \mbox{$3+1$}
components of the electric current density four-vector
(\ref{eq:CurrentEM}). The normal projection gives us the charge
density measured by Eulerian observers
\begin{equation}
\label{eq:rhoEM}
\rho_{{\rm e}} = - n_\mu j_{{\rm e}}^\mu
= \frac{iq}{2} \left( \Picc \vphi - \vphic \Pic \right) \; ,
\end{equation}
while the spatial projection is the electric current density measured
by Eulerian observers
\begin{equation}
\label{eq:jEM}
(j_{\rm e})^{i} = \gamma^{i}_\mu j_{{\rm e}}^\mu
= \frac{iq}{2}  \left( \vphic \chic ^i - \chicci \vphi \right) \; .
\end{equation}
It is also useful to express these quantities just in terms of the
scalar field quantities and the electromagnetic potentials:
\begin{eqnarray}
\label{eq:chrhofull}
\rho_{{\rm e}} &=& - q \: \mathbf{Im}(\Pi^\ast \vphi)
- q^2 \Phi \vphi^*\vphi , \\
\label{eq:chcurrfull}
(j_{{\rm e}})_i &=& + q \: \mathbf{Im}(\chi_i^\ast \vphi)
- q^2 a_i \vphi^*\vphi .
\end{eqnarray}

\subsection{Stress-Energy tensor}
\label{sec:stress}

The sources of the gravitational field are encoded in the
stress-energy tensor which is built from the contributions of the
electromagnetic and scalar fields
\begin{equation}
\label{eq:tmunuFull}
T^{\mu\nu} = {T_{\rm e}}^{\mu\nu} + {T_{\vphi}}^{\mu\nu} \; .
\end{equation}
This tensor appears in the evolution and constraint equations through
the 3+1 projections which correspond to the energy density $E$,
momentum density $J^i$, and stress tensor $S_{ij}$ measured by
Eulerian observers:
\begin{eqnarray}
{\mathcal E} &=& n^\mu n^\nu T_{\mu\nu} \; , \\
J^i &=& -\gamma^{i \mu} n^\nu T_{\mu\nu} \; , \\
S_{ij} &=& \gamma^\mu_i \gamma^\nu_j T_{\mu\nu} \; .
\end{eqnarray}

The contributions from the electromagnetic field can be written
explicitly as~\cite{Alcubierre:2009ij}
\begin{eqnarray}
\label{Edens}
{\cal E}_{\rm e} &:=& \frac{1}{8\pi} \left( E^2 + B^2 \right) \; , \\
\label{Poynt}
(J_{\rm e})^i &:=& \frac{1}{4\pi} \: {\epsilon^i}_{jk} E^j B^k \; , \\
\label{Stressten}
(S_{\rm e})_{ij} &:=& \frac{1}{8\pi} \left[ \gamma_{ij} \left(
E^2 + B^2 \right) \right. \nonumber \\
&-& \left. 2 \left( \rule{0mm}{3mm} E_i E_j + B_i B_j \right) \right] . 
\end{eqnarray}
In these expressions $E^2=\gamma_{ij}E^i E^j$ and $B^2=\gamma_{ij}B^i
B^j$.  In spherical symmetry, since there is no magnetic field, the
momentum density vanishes and we are left with the simple expressions
\begin{eqnarray}
\label{Edens2}
{\mathcal E}_{\rm e} &:=& \frac{1}{8\pi} \: \psi^4 A \left( E^r \right)^2 \; , \\
\label{Stressten2}
(S_{\rm e})^r_r &:=& -\frac{1}{8\pi} \: \psi^4 A \left( E^r \right)^2 \; , \\
(S_{\rm e})^\theta_\theta &:=& \frac{1}{8\pi} \: \psi^4 A \left( E^r \right)^2 \; .
\end{eqnarray}

For the scalar field, we start by writing the Lagrangian in terms of
the auxiliary variables $\Pic$ and $\chic$ as
\begin{equation}
L = - \frac{1}{2} \left( \chic^\ast_i \chic^i -\Pic^\ast \Pic
+ 2 V(|\varphi|^2) \right) \; .
\label{eq:lagC2}
\end{equation}
From this we can now calculate the stress-energy tensor and the
corresponding $3+1$ projections:
\begin{eqnarray}
\label{eq:rho}
{\mathcal E}_\vphi &=& \frac{1}{2}\left( \Pic^\ast \Pic
+ \chic_{i}^\ast \chi^i \right) + V \; , \\
\label{eq:j1}
(J_\vphi)^i &=&  - \frac{1}{2} \left( \Pic^\ast \chic^i
+ {\chic^{\ast i}} \Pic \right) \; , \\
\label{eq:Saux}
(S_\vphi)_{ij} &=& \frac{1}{2}\left( \chic^{\ast}_i \chic_j
+ \chic^{\ast}_j \chic_i \right) + \gamma_{ij} L \; ,
\end{eqnarray}
which in spherical symmetry reduces to
\begin{eqnarray}
\label{eq:rho2}
{\mathcal E}_\vphi &=& \frac{1}{2} \left( |\Pic|^2
+ \frac{|\chic|^2}{A\psi^4} \right) + V \; , \\
\label{eq:j12}
(J_\vphi)_r &=& - \frac{1}{2} \left( \Pic^\ast \chic
+ \chic^\ast \Pic \right) \; , \\
\label{eq:Saux2}
(S_\vphi)_r^r &=& \frac{1}{2}\left( |\Pic|^2
+ \frac{|\chic|^2}{A\psi^4} \right) - V \; , \\
\label{eq:Saux3}
(S_\vphi)_\theta^\theta &=& \frac{1}{2}\left( |\Pic|^2
- \frac{|\chic|^2}{A\psi^4} \right) - V \; .
\end{eqnarray}

\section{Initial data}
\label{sec:idata}

When looking for suitable initial data that represents a collapsing
charged scalar field it is not possible to assign freely all the
quantities since the Hamiltonian and momentum constraints
\begin{eqnarray}
\label{eq:ham}
^{(3)}R - K_{i j}K^{ij} + K^{2} &=& 16\pi {\mathcal E} \; , \\
\label{eq:mom}
D_{m} K^{ m }_i - D_i K &=& 8 \pi J_i \; ,
\end{eqnarray}
along with the electromagnetic
constraints~\eqref{eq:Max3}-\eqref{eq:Max4} have to be satisfied.
Notice that in spherical symmetry the momentum constraint has only one
non-trivial component, and the magnetic constraint becomes
trivial.

The first strong assumption we will make (apart from the spherical
symmetry) is that our initial data is momentarily stationary with zero
shift vector, which is equivalent to asking for the extrinsic
curvature $K_{ij}$ and the normal derivatives of scalar quantities to
vanish. This assumption, although not generic, is quite useful since
as long as we can make sure that the momentum density vanishes
initially the momentum constraint will be satisfied trivially.

We proceed further to simplify the Hamiltonian constraint by writing
the physical metric as $\gamma_{ij}=\psi^4 \hg_{ij}$, with $\hg_{ij}$
some known conformal metric. In terms of these quantities the
Hamiltonian constraint takes the form
\begin{equation}
\label{eq:AxisHam}
\frac{8}{\psi^5} \: \hnabla^2\psi - \hR + 16\pi {\mathcal E} = 0 \; ,
\end{equation}
with $\hR$ the Ricci scalar associated with the conformal metric.  For
a given conformal metric this is an elliptic equation for the
conformal factor $\psi$. This has to be solved simultaneously with the
Gauss constraint which can also be simplified by a conformal
rescaling. Defining the conformal electric field as \mbox{$\hat E^i=
  \psi^6 E ^i$}, the Gauss constraint takes the form
\begin{equation}
\label{eq:egauss}
\hnabla_i \hat E^i = 4 \pi \rho_e \psi^6 \; .
\end{equation}

A further assumption we can make is to take the conformal metric to be
flat.  By doing this the Laplacian and divergence are the usual ones
on flat space and the conformal Ricci scalar $\hR$ vanishes.  In
spherical symmetry, and with the matter content we are considering,
the Hamiltonian constraint takes the final form
\begin{eqnarray}
\label{eq:hamconf3}
\frac{1}{r^2} \: \p_r \left( r^2 \p_r \psi \right) 
+ 2\pi |\chic|^2 \psi \hspace{20mm} \nonumber \\ 
+ \left( \pi q^2 \Phi^2 | \vphi |^2 + 2\pi V\right)\psi^5
+ \frac{(\hat E^r)^2}{4 \psi^3} &=& 0 \; ,
\end{eqnarray}
while the Gauss constraint takes the form
\begin{equation}
\label{eq:gaussconf3}
\frac{1}{r^2} \: \p_r \left( r^2 \hat E^r \right) - 4 \pi \rho_e \psi^6  = 0 \; . 
\end{equation}

To solve these equations it is also necessary to impose boundary
conditions. The equation for the electric field $E^r$ is only first
order, so it is enough to specify the value of the field at the
origin, which due to the symmetry must be $E^r(r=0)=0$. The equation
for the conformal factor $\psi$, on the other hand, is second order so
we need to specify two boundary conditions.  At the origin we simply
ask for $\left. \partial_r \psi \right|_{r=0} = 0$, while far away we
ask for $\psi(r \rightarrow \infty)=1+C/r$ for some constant $C$,
which in turn implies that at the outer boundary $\partial_r \psi = (1
- \psi)/r$.

Below we will consider two distinct families of initial data, one
corresponding to a case with zero global charge, and a second one with
a non-zero global charge.

\subsection{Zero global charge}
\label{sec:zgc}

The simplest choice for the initial electromagnetic potentials is to
just set them equal to zero.  With this choice the charge
density~\eqref{eq:chrhofull} becomes proportional to $\Pi$, and the
current density~\eqref{eq:chcurrfull} proportional to $\chi$. We will
also assume that $\Pi=0$, so the charge density $\rho_e$ vanishes
initially and the spacetime has a vanishing total charge. The current
density $(j_e)_r$ does not vanishes as long as the product $\chi^\ast
\vphi$ has a non-zero imaginary part.  This in fact already excludes
the naive profile $\vphi = \vphi_0(r) e^{i\theta}$ for constant
argument $\theta$, which has as particular cases a purely real or
purely imaginary $\varphi$.  We will then consider initial data of the
form:
\begin{equation}
\varphi = f(r) + i g(r) \; ,
\end{equation}
with $f(r)$ and $g(r)$ real functions.  We then find that $(j_e)_r = i
q (g f' - f g')$.  So as long as $f$ and $g$ are distinct functions
that overlap is some region we will have a non zero current
density. In our simulations below we will take $f$ and $g$ as
gaussians centered at slightly different locations.

With these choices it is also easy to verify that the momentum
density~\eqref{eq:j12} vanishes, so the momentum constraints are
trivially satisfied.  That is, initially we have zero charge density
and zero momentum density (no energy flux), but non-zero current
density.  Physically this situation is analogous to a system with
identical initial distributions of particles of equal mass and
opposite charge, moving in opposite directions.

Since the initial charge density vanishes, the Gauss
constraint~\eqref{eq:egauss} implies that for regular initial data the
electric field must also vanish initially. We therefore need only to
solve for the conformal factor from the Hamiltonian constraint, which
now takes the simple form
\begin{equation}
\label{eq:eqhamconf2}
\frac{1}{r^2} \: \p_r \left(r^2 \p_r \psi \right) 
+ \pi |\chi|^2 \psi +2\pi V \psi^5 = 0 \; .
\end{equation}

In this specific case the Hamiltonian constraint takes exactly the
same form it would take if the scalar field was decoupled from the
electromagnetic field. It is also remarkable that
equation~\eqref{eq:eqhamconf2} does not depend on the value of the
fundamental charge of the scalar field $q$, so the initial data sets
obtained with this method can be evolved for arbitrary values of $q$.

\subsection{Net global charge}
\label{sec:ngc}

We are also interested in constructing initial configurations that
have non-zero total charge since we would like to study the case when
the initial configuration is such that $Q/M>1$. From equations
\eqref{eq:chrhofull} and \eqref{eq:chcurrfull} we see that this is
possible by making a few assumptions. First, since we are considering
a momentarily stationary configuration with no shift, we must still
have $\Pi=0$.  We can obtain configurations with non-vanishing charge
density and vanishing current density by setting $a_r=0$, $\Phi \neq
0$, and $\mathbf{Im}(\vphi)=0$.  This choice represents an initially
vanishing electric current with non-vanishing charge density given by
$\rho_{{\rm e}}= -q^2 \vphi^2 \Phi$.

With these choices the momentum density still vanishes since $\Pi=0$,
and both $\varphi$ and $\chi$ are purely real, so the momentum
constraints are again trivially satisfied.  Since the charge density
is non-zero, we now have to solve simultaneously the Hamiltonian and
Gauss constraints once we have chosen both $\varphi$ and $\Phi$.

Before solving the constraints we can look for some properties of
initial data with non-zero charge density by considering a simple
model where the charged matter is contained in a thin spherical shell
located at $r=r_0$. By Israel's theorem~\cite{Heusler96} the exterior
region of the spacetime corresponds to the Reissner-N\"ordstrom (RN)
solution~\cite{Reissner16,Nordstrom18}, which for the initial slice
can be written as
\begin{equation}
\label{eq:RNpsi}
\psi = \left[ \left( 1 + \frac{M}{2r} \right)^2 
- \left( \frac{Q}{2r} \right)^2 \right]^{1/2} \; ,
\end{equation}
and
\begin{equation}
\label{eq:RNE}
\hat E^r = \frac{Q}{r^2} \; .
\end{equation}
It is straightforward to verify that these functions are solutions to
the constraints \eqref{eq:hamconf3}-\eqref{eq:gaussconf3} when the
scalar field vanishes.

The solution in the interior of the thin shell is trivial since we are
looking for regular initial data.  A flat metric with vanishing
electric field is the only solution that is regular at $r=0$, and is
characterized only by a constant conformal factor $\psi_0$.  These
solutions can be matched at $r_0$ to give the spatial metric of the
initial slice, and this determines the value of $\psi_0$. The RN
geometry in this foliation is known to have a trapped surface at
\mbox{$r=\sqrt{M^2-Q^2}/2$} which only exists if $M>|Q|$.  If the
initial shell has a radius smaller that this value, then there is an
initial black hole in the spacetime and all the matter is enclosed by
the horizon.  This type of solution is of little interest since
distant observers would not distinguish it from a stationary RN black
hole. On the other hand, when $|Q|>M$ there exists a pathological
behavior in these coordinates, because at the finite radius $r_p =
(|Q|-M)/2$ the conformal factor $\psi$ becomes zero. This corresponds
to the place where the singularity of the over-extremal RN solution is
mapped in these coordinates, and the region $r<r_p$ has no physical
relevance.

The analysis of this simplified model shows that we should be cautious
when solving the constraint equations for large initial charge
densities.  In that case, if we ask for the matter to be tightly
distributed close to the origin in a region of radius $R$, one could
find that $R<r_p$ in which case the initial data will not represent an
asymptotically flat spacetime. Numerically, we find that as we
construct configurations of localized shells with higher $|Q|/M$
eventually the algorithm fails as a consequence of the conformal
factor becoming zero outside the scalar field shell.  This was not a
problem for the case of vanishing initial charge density presented
before, since in that case the solution approaches the Schwarzschild
metric outside the matter which does not present such pathologies.

\section{Analysis tools and numerical code}
\label{sec:numerical}

For the numerical simulations discussed here we integrate the
equations obtained in Section~\ref{sec:EMKG}, along with the BSSN
equations, using a finite difference scheme. The code uses a method of
lines with second or fourth order spatial differences, along with
either three-step iterative Crank-Nicholson or fourth order
Runge-Kutta time integrators. This code has been previously tested
with real scalar fields, and has been used in the context of
scalar-tensor theories of gravity with minimal
modifications~\cite{Alcubierre:2010ea,Ruiz:2012jt}.

The evolutions presented below were performed on three different
resolutions to rule out discretization effects, namely $\Delta r =
0.02,0.01,0.005$, with 3000, 6000 and 12000 grid points respectively.

\subsection{Gauge choice}
\label{sec:gauge}

For our simulations we choose for simplicity a vanishing shift,
whereas for the lapse function we choose the maximal slicing condition
$K=\p_t K=0$, which leads to an elliptic equation for the lapse
function $\alpha$ that takes the following form in spherical symmetry
\begin{eqnarray}
\label{eq:maximalSlicing}
\p^2_r \alpha &+& \left( \frac{2}{r} - \frac{\p_r A}{2A}
+ \frac{\p_r B}{B} + 2 \p_r \ln \psi \right) \p_r \alpha \nonumber \\
&=& \alpha A \psi^4 \left[ K^2
+ 4 \pi \left( {\mathcal E} + S \right) \right] \; ,
\end{eqnarray}
with $K^2= K_{ij} K^{ij}$, and $S$ the trace of $S_{ij}$.  Since in
spherical symmetry this is a linear ordinary differential equation for
$\alpha$, it can be solved numerically at each time step by matrix
inversion without increasing significantly the computational costs of
the whole scheme.

\subsection{Regularization}
\label{sec:reg}

The origin in spherical coordinates is a singular point and its
inclusion in the numerical domain may be problematic due to
instabilities that arise from numerical errors. The problem comes from
the fact that the coordinate transformation induces terms on the
evolution equations that go as $1/r$ and seem ill behaved.  These
divergences cancel analytically, but this may not be the case when we
consider the numerical error.

To avoid this problem one needs to take two steps.  First, we use a
grid that staggers the origin and enforce the parity conditions that
must be satisfied by the different functions in order to be regular
depending on their tensorial character. Also, special care needs to be
taken with the evolution equations for the metric and extrinsic
curvature since they contain terms that are not manifestly regular at
$r=0$.  One then needs to define certain combinations of quantities as
new independent variables and evolve them separately
(see~\cite{Alcubierre04a,Alcubierre:2010is} for details).

One should mention the fact that it has recently been pointed out by
Montero and Cordero-Carrion~\cite{Montero:2012yr} that one can avoid
the need of a special regularization algorithm if one uses a partially
implicit Runge-Kutta method for the time integration.  In fact we have
found that with the numerical methods used in our code, which are
fully explicit, one can also obtain stable and convergent evolutions
without regularization, but at he price of having the numerical errors
close to the origin increase considerably.  Using the regularization
algorithm increases the accuracy significantly at no serious extra
computational cost.

\subsection{Mass and Charge content of the spacetime}
\label{sec:chconservation}

A convenient way to determine the mass of a localized distribution
that has spherical symmetry is by considering the radial dependence of
the metric components. When expressed in terms of the areal radius
$R$, in which the area of a sphere is $4\pi R^2$, the radial component
of the metric behaves as
\begin{equation}
\label{eq:MassSCH}
g_{RR}=\left( 1- \frac{2M(R)}{R} \right)^{-1} \; .
\end{equation}
After a little algebra we can write the function $M(R)$ in terms of
our original radial coordinate $r$ as
\begin{equation}
\label{eq:MassSCH2}
M(r) = \frac{r \psi^2 B^{1/2}}{2}
\left[ 1 - \frac{B}{A} \left( 1
+ r \frac{\p_r B}{2B} + 2 r \frac{\p_r\psi}{\psi} \right)^2 \right] .
\hspace{7mm}
\end{equation}
This function may be identified with the total mass outside the matter
sources, since it attains the value of the ADM mass in the vacuum
region. In our case, since the electric field extends to infinity, the
value of the mass function is always less than the total ADM
mass, but approaches it quickly since the electric field decays as $1/r$.

For the electric charge, all information is contained in Maxwell's
equations. We can define the total charge contained in a region of a
constant $t$ hypersurface by integrating the charge density $\rho_e$
over that region.  Since we are working on spherically symmetric
slices it is convenient to define the charge enclosed inside a sphere
of radius $r$ as
\begin{equation}
\label{eq:charge}
Q(r) = \int_{S(r)}{\rho_e \: dV} .
\end{equation}
Taking the limit when $r\rightarrow\infty$ we get the total charge of
the spacetime, which is a conserved quantity as a consequence of the
4-current $j_e^\mu$ satisfying the continuity equation $\nabla_\mu
j_e^\mu=0$. By using Gauss law and the divergence theorem, this
expression can be converted to a boundary integral as follows
\begin{eqnarray}
\label{eq:charge2}
Q(r) &=& \frac{1}{4\pi}\int_{S(r)}{D_a E^a dV} \nonumber{}\\
   &=&\frac{1}{4\pi}\oint_{\p S(r)}{{\hat r}_a E^a dS} \; ,
\end{eqnarray}
where we have assumed that the interior of the sphere is regular, and
where $\hat r$ is the outward pointing unit normal vector to the
sphere. In spherical symmetry the angular dependency can be integrated
immediately yielding
\begin{equation}
\label{eq:charge3}
Q(r) = r^2 \psi^6 \sqrt{A}B E^r \; .
\end{equation}

Since in our simulations the numerical domain is finite we can only
consider the charge enclosed by spheres of finite radius, which won't
be conserved since there may be scalar field that is scattered to
infinity carrying away electric charge with it beyond the
computational domain. It is possible, however, to track the rate of
change of the enclosed charge by using the continuity equation. After
some algebra we find
\begin{equation}
\label{eq:dq}
\frac{dQ(r)}{dt} = \int_{S(r)}{ D_i \left( \rho_e \beta^i
- \alpha \: {}^{(3)}j_e^i \right) } dV \; ,
\end{equation}
where we have kept the shift vector dependence for generality.  Again,
this is the integral of a divergence which can be transformed into a
boundary integral, and the angular dependence can be immediately
integrated yielding
\begin{equation}
\label{eq:dq2}
\frac{dQ(r)}{dt} = 4 \pi r^2 \psi^6 \sqrt{A}B 
\left( \rho_e \beta^r - \alpha \: {}^{(3)}j_e^r \right)  \; .
\end{equation}
The above equation can be integrated in time to find the change of the
enclosed charge at fixed coordinate radius $r$ after a time $T$
\begin{equation}
\label{eq:dqint}
\Delta Q(r,T) = 4 \pi r^2 \int_0^T {\psi^6 \sqrt{A} B 
\left( \rho_e \beta^r - \alpha {}^{(3)}j_e^r \right) dt} \: .
\end{equation}

\subsection{Horizons and irreducible mass}
\label{sec:ah}

Although the only formal way to identify the presence of a black hole
in the spacetime is by identifying its event horizon, one can learn a
lot about the dynamics of the collapse from the apparent
horizons. Since the strong energy condition holds for charged scalar
fields, the singularity theorems~\cite{Hawking73} imply that the
development of trapped surfaces will lead to a singularity in the
spacetime. Also, once the final black hole has settled down the
apparent horizon will coincide with the event horizon.  For a
stationary black hole the area $A_{\rm H}$ of its event horizon is
related to the so-called irreducible mass by
\begin{equation}
\label{BHmass1}
M_{\rm irr}= \sqrt{ \frac{A_{\rm H}}{16\pi} } \; .
\end{equation}
The irreducible mass gets this name because it is the minimum mass value
a black hole can attain for a fixed value of the area. The actual mass
of a non-rotating charged black hole is related to this quantity by
\begin{equation}
\label{eq:bhmass2}
M_{\rm H} = M_{\rm irr} + \frac{Q_{\rm H}^2}{4M_{\rm irr}} \; ,
\end{equation}
where the horizon charge $Q_{\rm H}$ is evaluated at the black hole surface.

\subsection{Characteristic adapted boundary conditions}
\label{sec:boundary}

The finiteness of the numerical domain must be dealt with carefully
since the boundary conditions imposed may introduce errors that affect
the results of the simulations. Usually one imposes outgoing wave
boundary conditions to avoid spurious reflections, but it turns out
that such conditions are not compatible in general with the evolution
equations. In the case of General Relativity the evolution equations
preserve the constraints but generic boundary conditions fail to do
so.  We will deal elsewhere with the problem of applying constraint
preserving boundary conditions for the BSSN formulation in spherical
symmetry~\cite{Alcubierre:2013prep}, while here we will just discuss
the issue of how to apply boundary conditions to the scalar and
electromagnetic fields.

We start by looking at the characteristic structure of the evolution
equations in order to find the ingoing and outgoing eigenfields. These
eigenfields are reconstructed at the boundary after all the dynamical
variables have been updated all the way to the boundary (using
one-sided differences at the boundary itself). We then apply suitable
boundary conditions to the incoming fields while leaving the outgoing
fields unchanged. After this we reconstruct the original dynamical
fields at the boundary.

In order to find the eigenfields associated with the scalar field, we
start from the fact the evolution system up to principal part has the
form
\begin{eqnarray}
\partial_t \vphi - \beta \partial_r  \vphi &\sim& 0 \; , \\
\partial_t \chi - \beta \partial_r  \chi - \alpha \partial_r \Pi
&\sim& 0 \; , \\
\partial_t \Pi - \beta \partial_r \Pi - \frac{\alpha}{A\psi^4} \:
\partial_r \chi &\sim& 0 \; .
\end{eqnarray}
Diagonalizing this system we find the following eigenfields $\omega$
and corresponding eigenspeeds $\lambda$:
\begin{eqnarray}
\label{eq:eigenmode0}
\omega^0_\vphi = \vphi \; , &\quad& \lambda^0_\vphi = - \beta \; , \\
\label{eq:eigenmodepm}
\omega^\pm_\vphi = \Pi \mp \frac{\chi}{\sqrt{A}\psi^2} \; , &\quad&
\lambda^\pm_\vphi = - \beta \pm \frac{\alpha}{\sqrt{A}\psi^2} \; . \hspace{5mm}
\end{eqnarray}
Notice that the characteristic speed $\lambda_0$ corresponds to
propagation along the normal direction to the hypersurfaces, while the
speeds $\lambda^\pm$ correspond to propagation along the light cone.

We now assume that far from the sources the scalar field behaves as a
spherical wave of the form \mbox{$\vphi=f(r-\lambda^+ t)/r$}.  This
can be easily shown to imply that the incoming eigenfield must have
the form
\begin{equation}
\label{eq:bcscalareigen}
\omega^-_\vphi = - \frac{\vphi}{r \sqrt{A} \psi^2} \; .
\end{equation}
Notice that, because of the $1/r$ decay of the spherical wave, the
incoming field is not zero as one could naively expect.  Setting
$\omega^-_\vphi=0$ at the boundary in fact results in large
reflections.  In practice, we evolve both $\Pi$ and $\chi$ all the way
to the boundary using one-sided derivatives, and use their values to
construct the outgoing field $\omega^+_\vphi$ at the boundary.  We
then calculate $\omega^-_\vphi$ from equation~\eqref{eq:bcscalareigen}
above, and finally reconstruct $\Pi$ and $\chi$ from the values
of $\omega^+_\vphi$ and $\omega^-_\vphi$.

The evolution equations for the electric field turn out to be
identical to those of the scalar field up to principal part.  One only
needs to make the identifications \mbox{$\Pi \rightarrow \Phi$}, $\chi
\rightarrow -a_r$, $\vphi \rightarrow E^r$. In this case the
eigenfields and corresponding speeds become
\begin{eqnarray}
\omega^0_E = E^r \; , &\quad& \lambda^0_E = - \beta \; , \\
\omega^\pm_E = \Phi \pm \frac{a_r}{\sqrt{A}\psi^2} \; , &\quad&
\lambda^\pm_E = - \beta \pm \frac{\alpha}{\sqrt{A}\psi^2} \; .
\end{eqnarray}
However, the analogy with the scalar field system ends here since the
electromagnetic potentials do not relate to the electric field as its
normal and tangential derivatives. Since in absence of a shift vector
the electric field evolves trivially up to principal part we just
update it using the values of the source terms on the boundary.

For the electromagnetic potentials we are in principle free to specify
the incoming eigenfield at the boundary.  However, just as it happened
with the scalar field, if we want to avoid large reflections at the
boundary the incoming field should not be chosen to be equal to zero.
But now we face a problem: in the case of the scalar field we could
model $\vphi$ as an outgoing spherical wave, and from that deduce the
form of $\omega^-_\vphi$ at the boundary.  But for the electric field
we can't do the same since, as we have already mentioned, the
electromagnetic potentials are not the time and space derivatives of
some other field.  Because of this we will simply model the incoming
field $\omega^-_E$ itself as a spherical outgoing wave at the
boundary, and consequently we will ask for it to satisfy an advection
equation of the form
\begin{equation}
\label{eq:outgoingrad}
\p_t \omega^-_E + \lambda \: \p_r \omega^-_E + \lambda \frac{\omega^-_E}{r} = 0 \; .
\end{equation}
This equation is evolved at the boundary to obtain the value of the
incoming field on the new time-step, and from this we reconstruct the
electromagnetic potentials at the boundary.

\section{Results of our numerical simulations}
\label{sec:results}

We will now discuss the results of some of our numerical simulations.
We will do it first for the configurations with zero global charge,
and after that for the configurations with non-zero initial charge.

\subsection{Globally uncharged configurations}
\label{sec:resultsuncharged}

For the first set of simulations we used the initial data
corresponding to scalar field configurations with vanishing global
charge. As discussed before, in this case we assume that the
electromagnetic potentials $\Phi$ and $a_r$ vanish initially.  We
choose a Gaussian profile for the scalar field that represents a shell
of matter with non-zero initial current density.  Explicitly, at $t=0$
we choose the following gaussian profiles for the real and imaginary
parts for the scalar field:
\begin{eqnarray}
{\rm Re}{(\vphi)} &=& \vphi_0 \left[ e^{-(r-r_R)^2/\sigma^2}
+ e^{-(r+r_R)^2/\sigma^2} \right] \: , \\
{\rm Im}{(\vphi)} &=& \vphi_0 \left[ e^{-(r-r_I)^2/\sigma^2}
+ e^{-(r+r_I)^2/\sigma^2} \right] \: ,
\end{eqnarray}
where we use the same amplitude $\vphi_0$ and width $\sigma$ for both
the real and imaginary part, but with the pulses centered on slightly
different places, $r_R \neq r_I$, in order to have a non-vanishing
electric current.  We also sum the mirror image of the Gaussian to
ensure that the scalar field is well behaved at the origin.

For all the simulations shown here we have chosen $r_R=5.0$, $r_I=5.1$
and $\sigma=1.0$, and we solve the Hamiltonian constraint for
different values of $\vphi_0$. We performed simulations for a two
parametric set of configurations, with the fundamental charge $q$
ranging from 0 to 8, and the initial amplitude of the pulses $\vphi_0$
ranging from $0.03$ to $0.1$.

\vspace{5mm}

\begin{figure}[htbp]
\centering
\includegraphics[width=9cm]{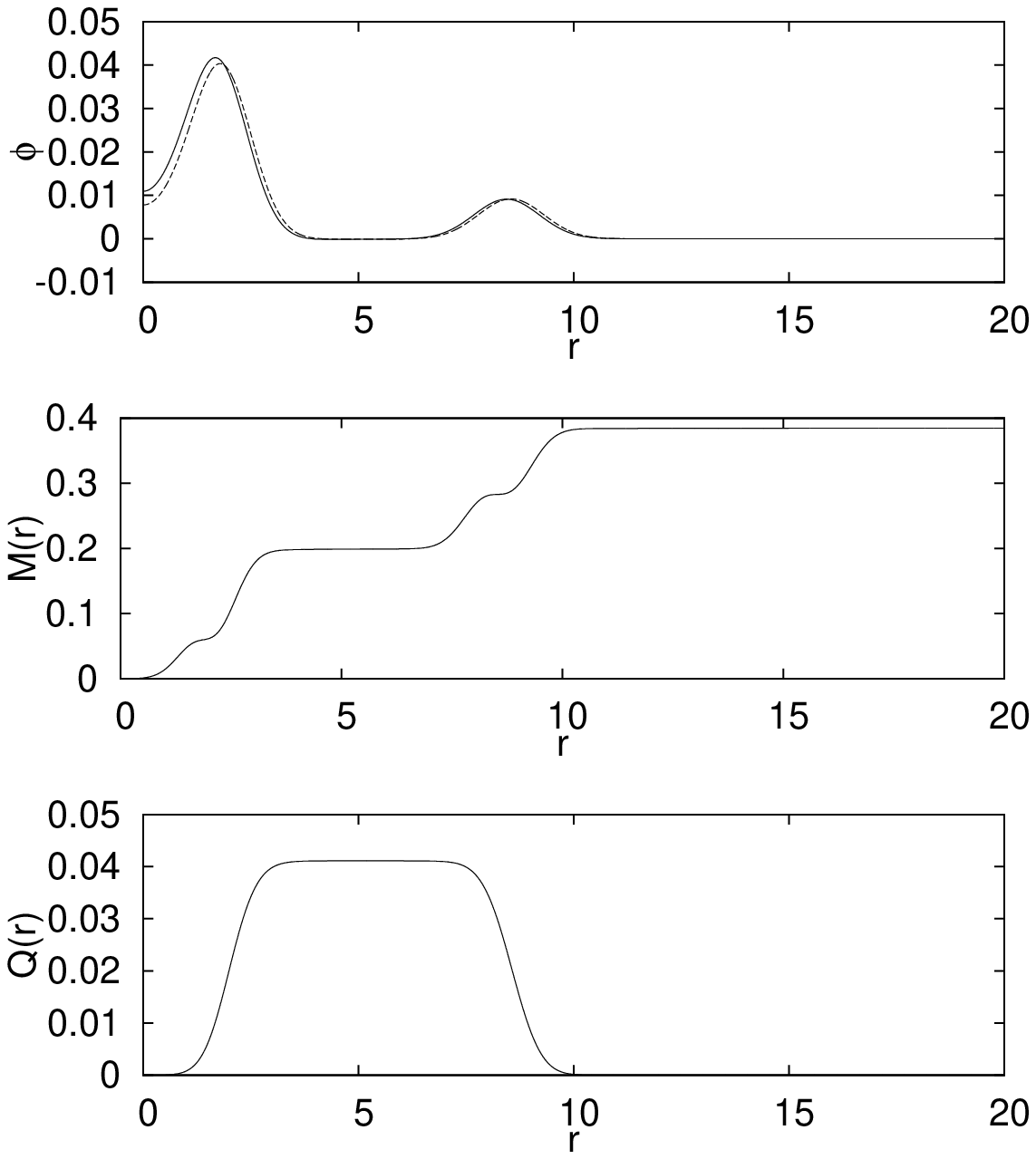}
\caption{Plot of the early stage of the evolution \mbox{($t=4$)} of a
  configuration with zero initial charge density and
  \mbox{$\vphi_0=0.03$}, \mbox{$q=2.0$}.  The top panel shows the
  scalar field (with the solid and dotted lines corresponding to the
  real and the imaginary parts), while the lower two panels show the
  integrated mass $M(r)$ and charge $Q(r)$.}
\label{fig:NQEarly}
\end{figure}

The simulations with low values of $\vphi_0$ behave as expected: The
scalar field propagates and is eventually dispersed to infinity
leaving behind flat spacetime. As an example, we consider the
evolution for the case with $\vphi_0=0.03$ and $q=2.0$. Initially the
pulse separates into incoming and outgoing parts, and a non-zero
charge density quickly develops.  This can be seen in
Figure~\ref{fig:NQEarly}, which shows a snapshot of the evolution at
$t=4$.  The top panel shows the scalar field, while the lower two
panels show the integrated mass $M(r)$ and charge $Q(r)$.

\begin{figure}[htbp]
\centering
\includegraphics[width=9cm]{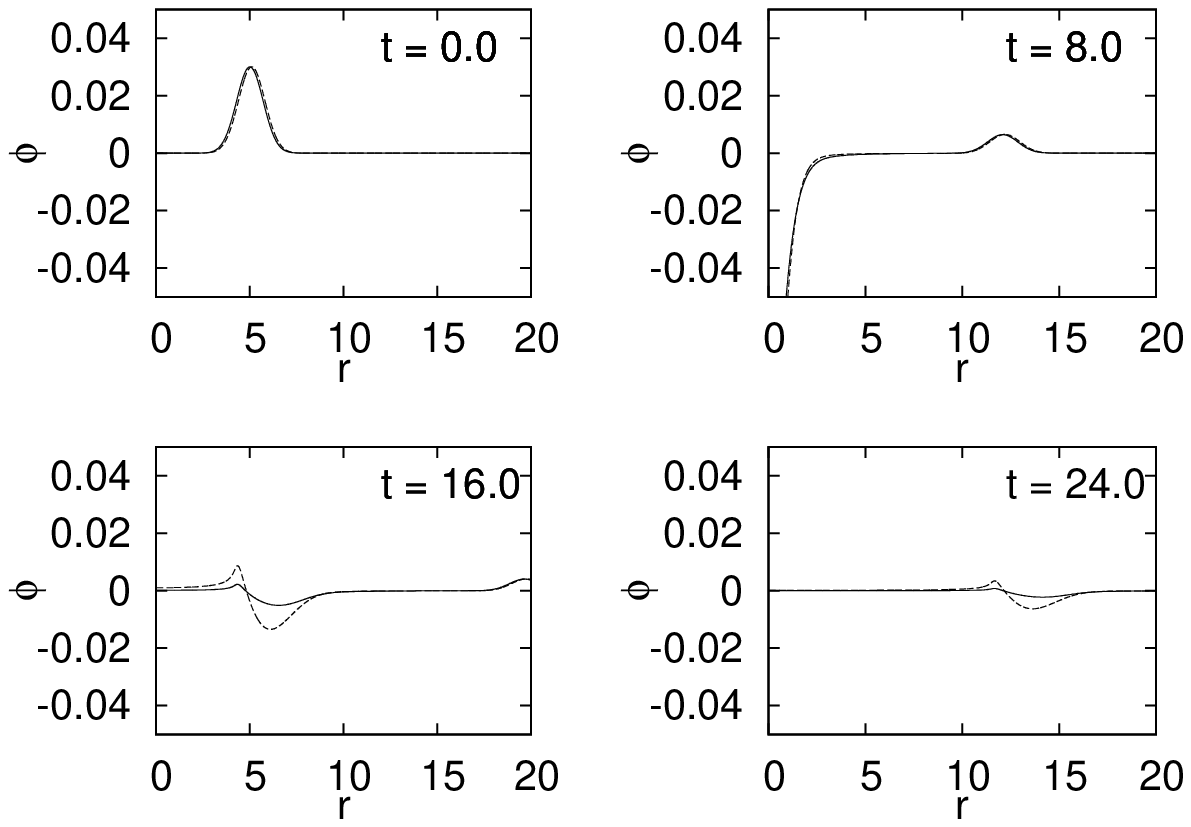}
\caption{Evolution of the scalar field for a configuration with zero
  initial charge density and \mbox{$\vphi_0=0.03$},
  \mbox{$q=2.0$}. The solid and dotted lines correspond to the real
  and the imaginary parts respectively.}
\label{fig:NQsnapshotsScalar}
\end{figure}

\begin{figure}[htbp]
\centering
\includegraphics[width=9cm]{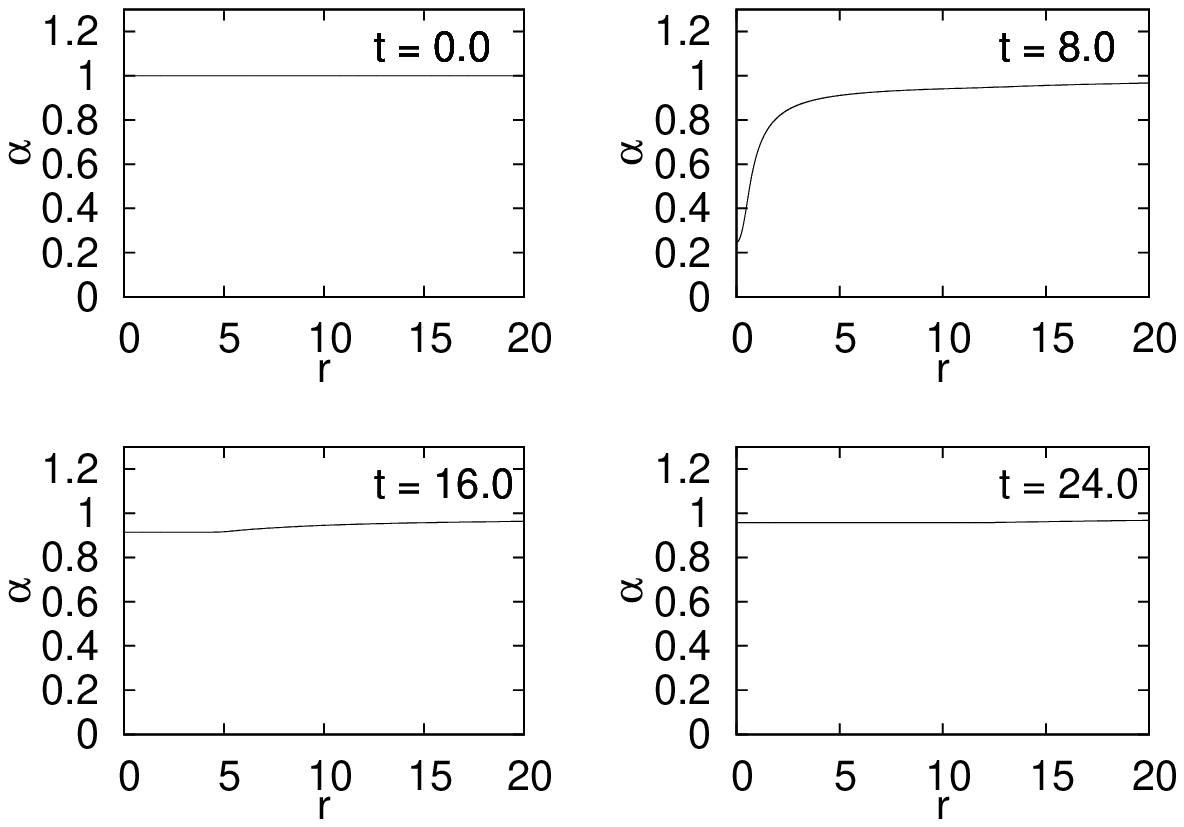}
\caption{Evolution of the lapse function $\alpha$ for a configuration
  with zero initial charge density and \mbox{$\vphi_0=0.03$},
  \mbox{$q=2.0$}.}
\label{fig:NQsnapshotsAlpha}
\end{figure}

\begin{figure}[htbp]
\centering
\includegraphics[width=9cm]{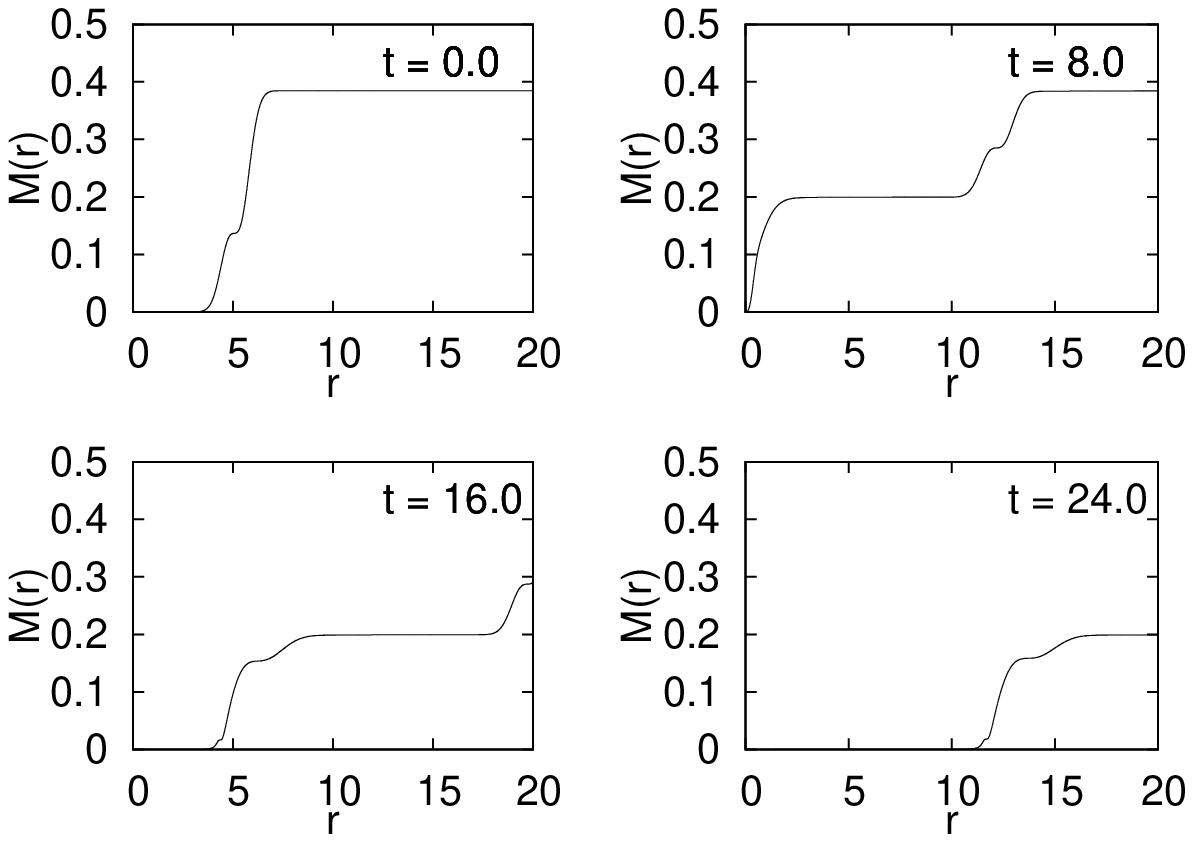}
\caption{Evolution of the integrated mass $M(r)$ for a configuration
  with zero initial charge density and \mbox{$\vphi_0=0.03$},
  \mbox{$q=2.0$}.}
\label{fig:NQMass}
\end{figure}

\begin{figure}[htbp]
\centering
\includegraphics[width=9cm]{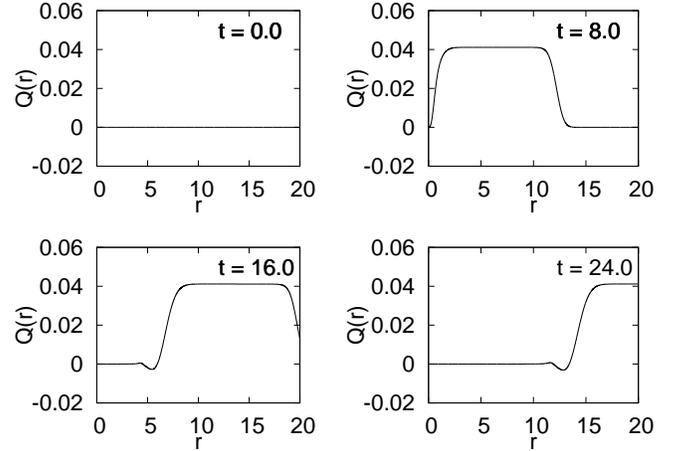}
\caption{Evolution of the integrated charge $Q(r)$ for a configuration
  with zero initial charge density and \mbox{$\vphi_0=0.03$},
  \mbox{$q=2.0$}. }
\label{fig:NQCharge}
\end{figure}

Figures~\ref{fig:NQsnapshotsScalar} and~\ref{fig:NQsnapshotsAlpha}
show the evolution of the scalar field and lapse function
respectively.  We note that at $t \sim 8$ the incoming pulse reaches
the origin and the lapse function decreases significantly
there. However, in this case the self-gravity of the scalar field is
not strong enough to produce a collapse, and the field later disperses
away to infinity while the lapse slowly returns to its flat space
value.
  Figures~\ref{fig:NQMass} and~\ref{fig:NQCharge} show the
evolution of the integrated mass and charge for this simulation.

\begin{figure}[htbp]
  \centering
  \includegraphics[width=8.5cm]{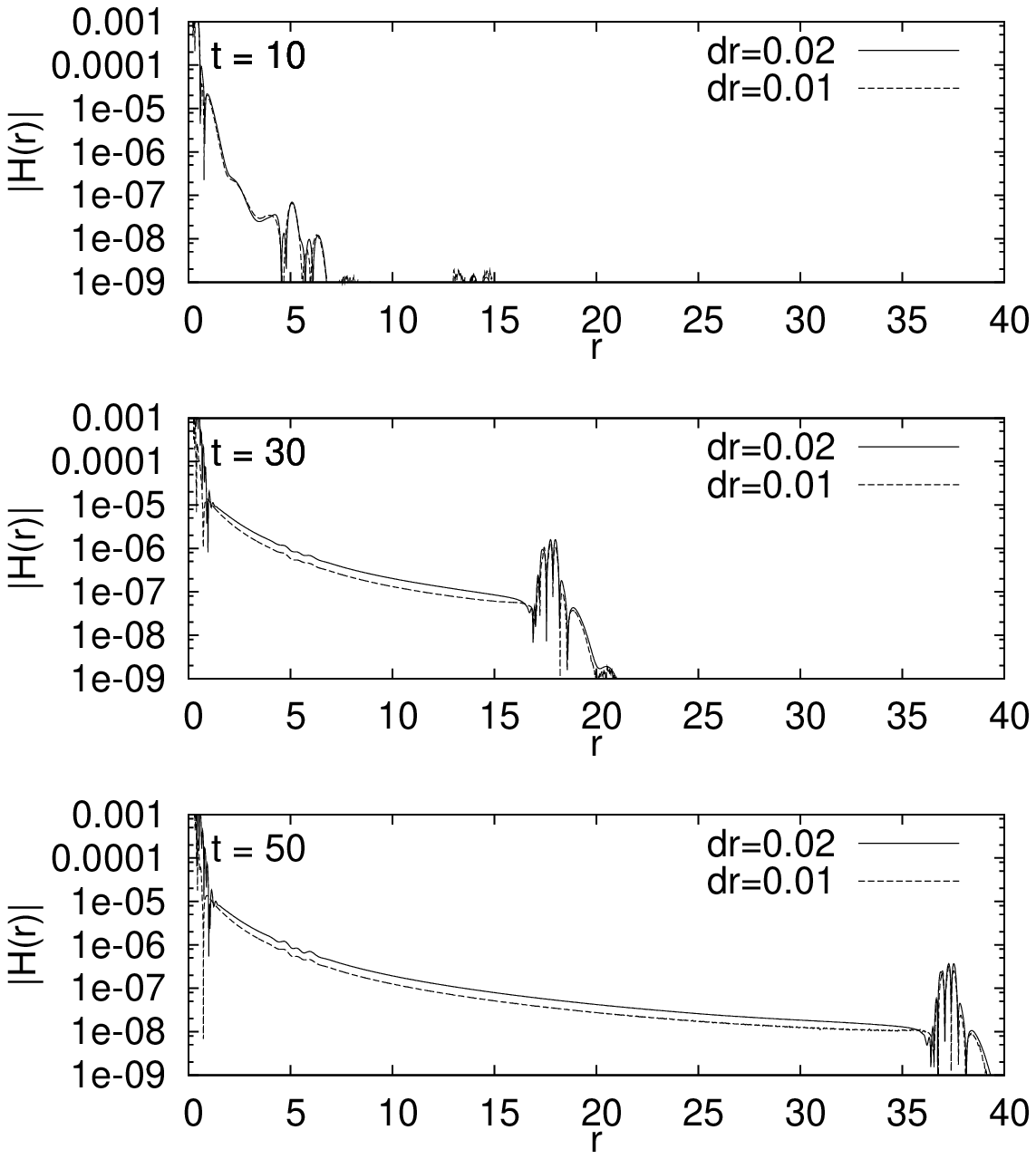}
  \caption{Plot of the absolute value of the Hamiltonian constraint
    residual evaluated throughout the numerical domain at different
    times for two different resolutions, for a configuration with zero
    initial charge density and \mbox{$\vphi_0=0.03$},
    \mbox{$q=2.0$}. The higher resolution has been multiplied by $16$
    showing the expected fourth order convergence.}
  \label{fig:uHAM}
\end{figure}

Convergence is verified by analyzing the constraint residuals that
arise from the discretization scheme.  Figure~\ref{fig:uHAM} shows the
absolute value of the Hamiltonian constraint at different times for
two different resolutions, showing that the error scales with
resolution consistently with the expected order of the discretization
(fourth order in this case).

\vspace{5mm}

\begin{figure}[ht]
\centering
\includegraphics[width=9cm]{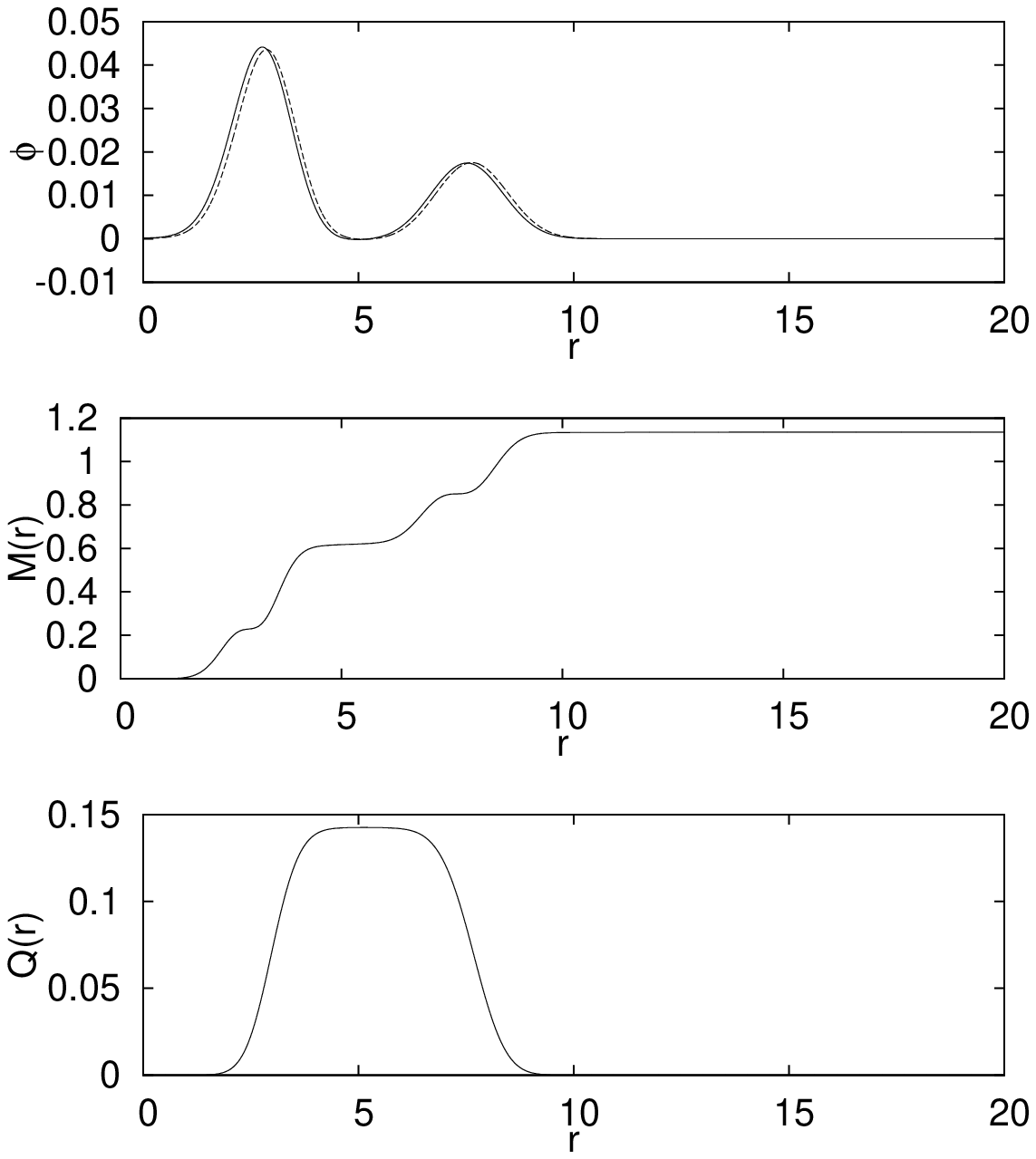}
\caption{Plot of the early stage of the evolution \mbox{($t=4$)} of a
  configuration with zero initial charge density and
  \mbox{$\vphi_0=0.05$}, \mbox{$q=2.0$}.  The top panel shows the
  scalar field (with the solid and dotted lines corresponding to the
  real and the imaginary parts), while the lower two panels show the
  integrated mass $M(r)$ and charge $Q(r)$.}
\label{fig:NQEarlyCol}
\end{figure}

\begin{figure}[htbp]
\centering
\includegraphics[width=9cm]{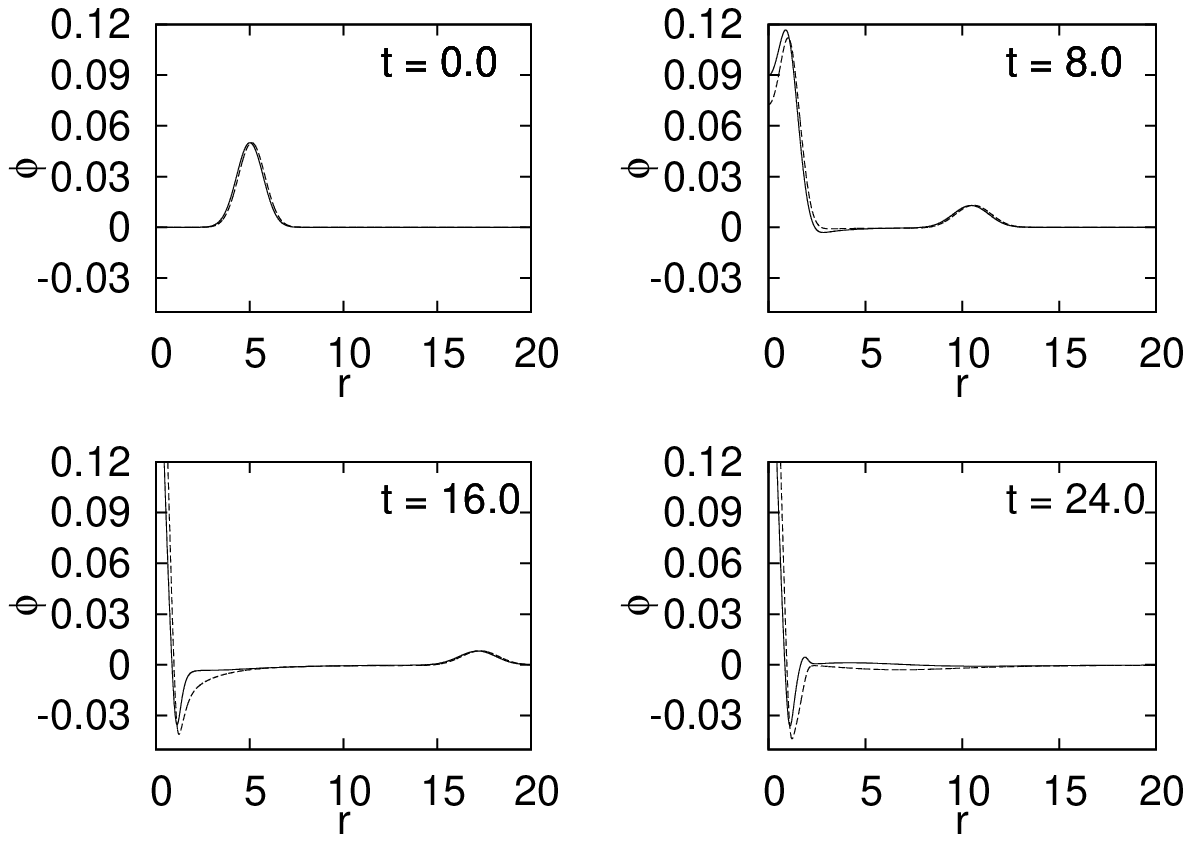}
\caption{Evolution of the scalar field for a configuration with
  zero initial charge density and \mbox{$\vphi_0=0.05$},
  \mbox{$q=2.0$}.  The solid and dotted lines correspond to the real
  and the imaginary parts respectively.}
\label{fig:NQsnapshotsScalarCol}
\end{figure}

\begin{figure}[htbp]
\centering
\includegraphics[width=9cm]{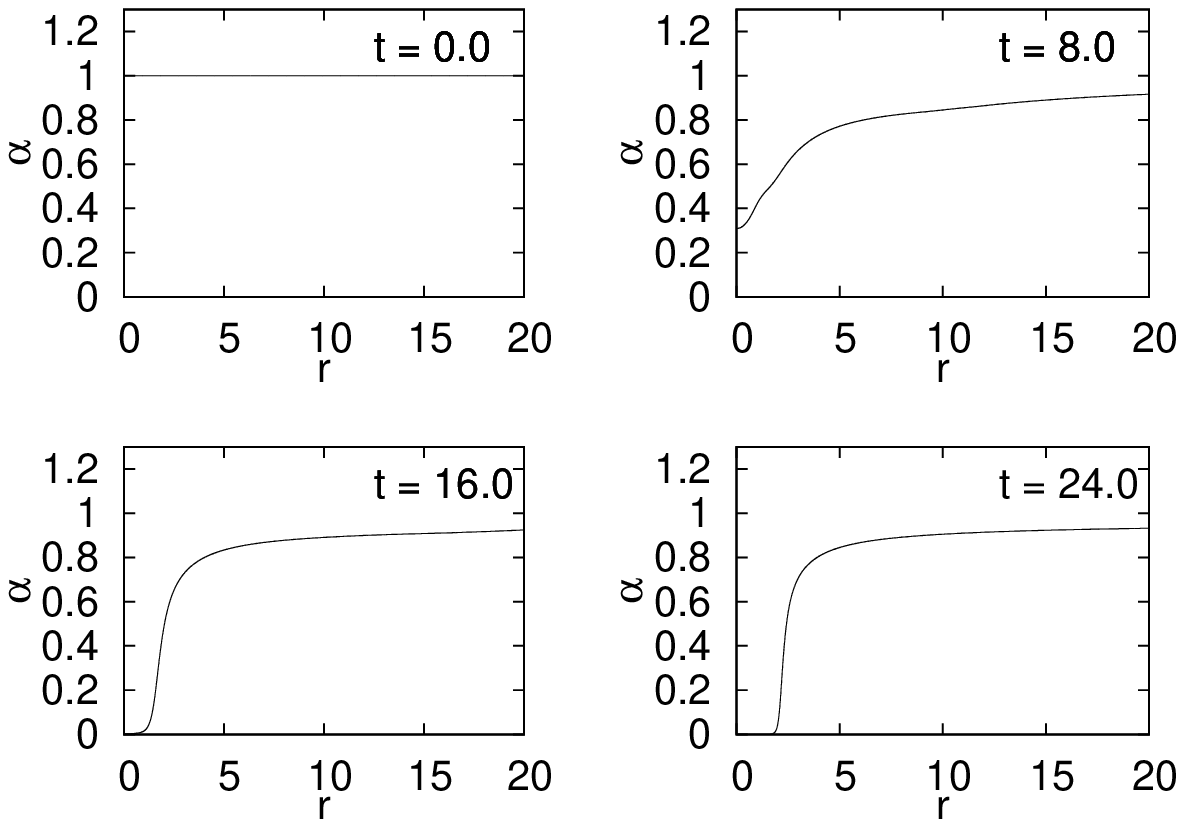}
\caption{Evolution of the lapse $\alpha$ for a configuration with zero
  initial charge density and \mbox{$\vphi_0=0.05$}, \mbox{$q=2.0$}.}
\label{fig:NQsnapshotsAlphaCol}
\end{figure}

\begin{figure}[htbp]
\centering
\includegraphics[width=9cm]{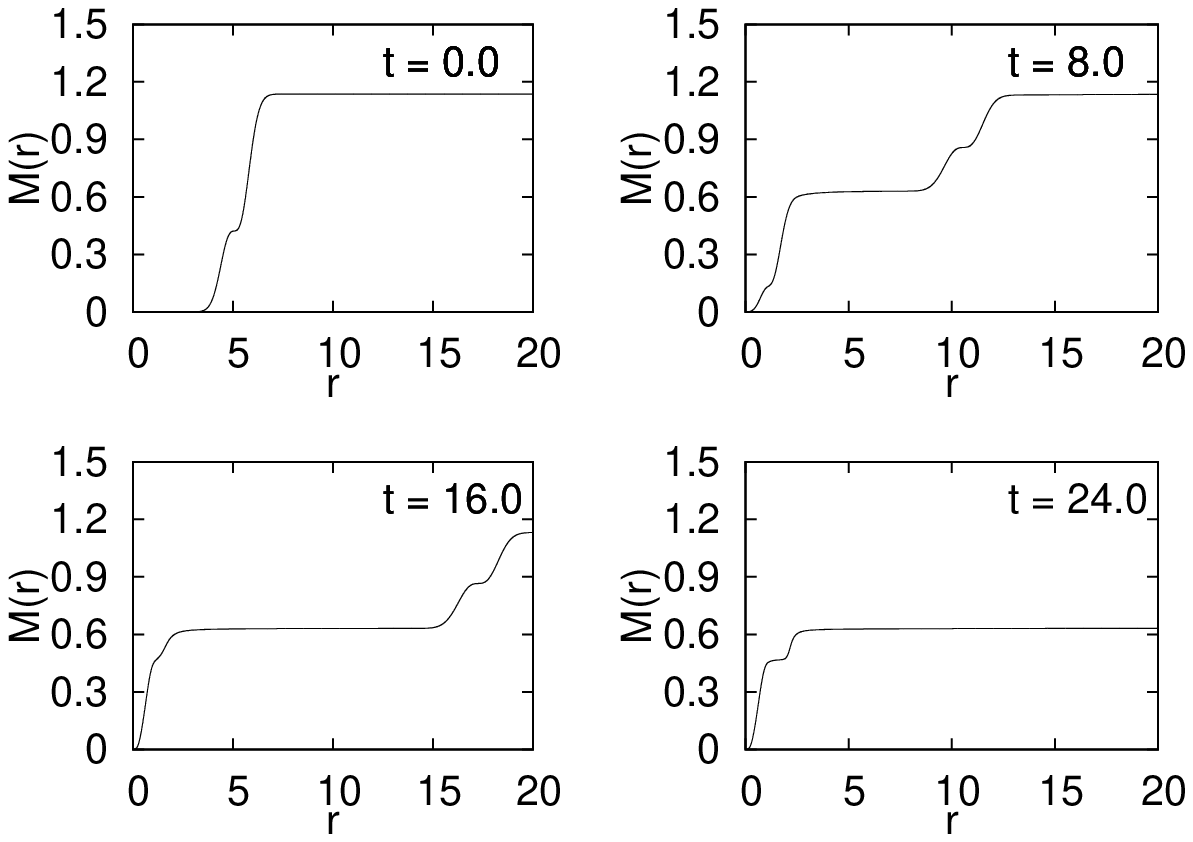}
\caption{Evolution of the integrated mass $M(r)$ for a configuration
  with zero initial charge density and \mbox{$\vphi_0=0.05$},
  \mbox{$q=2.0$}. }
\label{fig:NQMassCol}
\end{figure}

\begin{figure}[htbp]
\centering
\includegraphics[width=9cm]{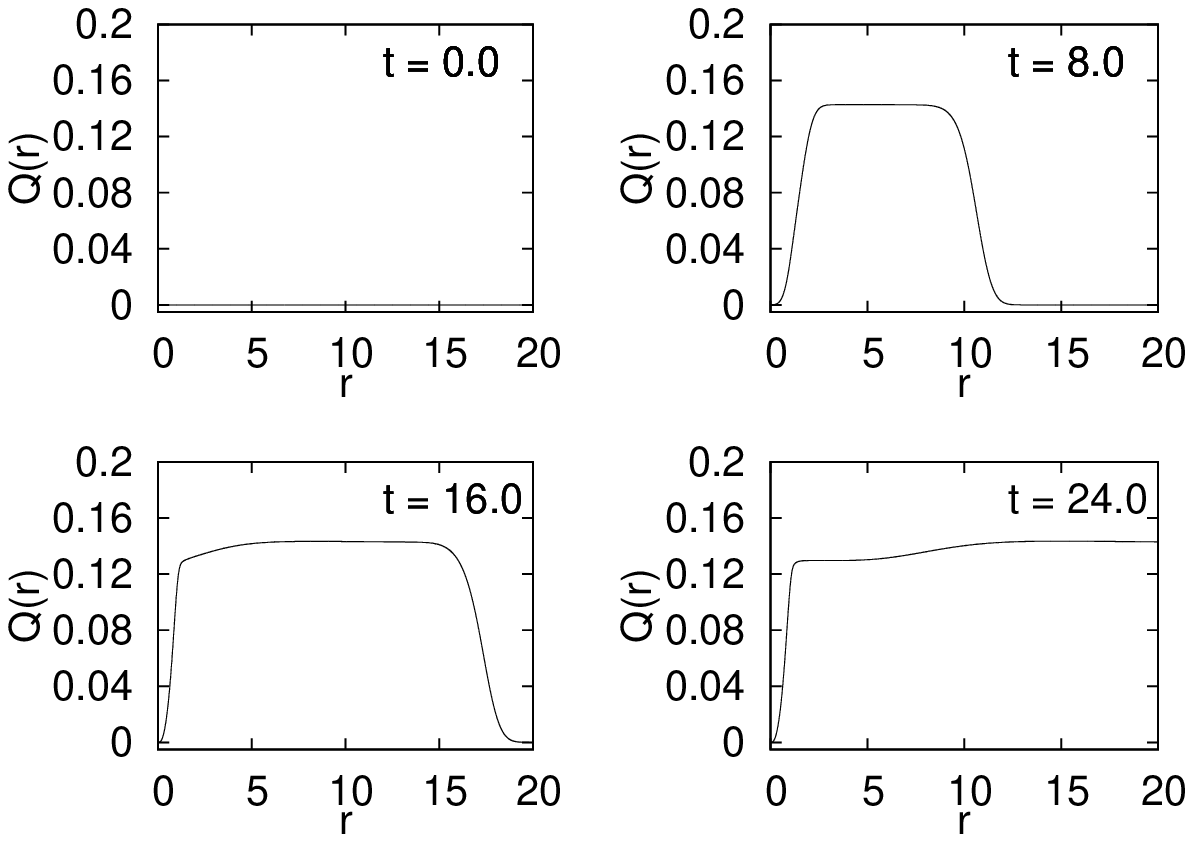}
\caption{Evolution of the integrated charge $Q(r)$ for a configuration
  with zero initial charge density and \mbox{$\vphi_0=0.05$},
  \mbox{$q=2.0$}.}
\label{fig:NQChargeCol}
\end{figure}

\begin{figure}[htbp]
\centering
\includegraphics[width=9cm]{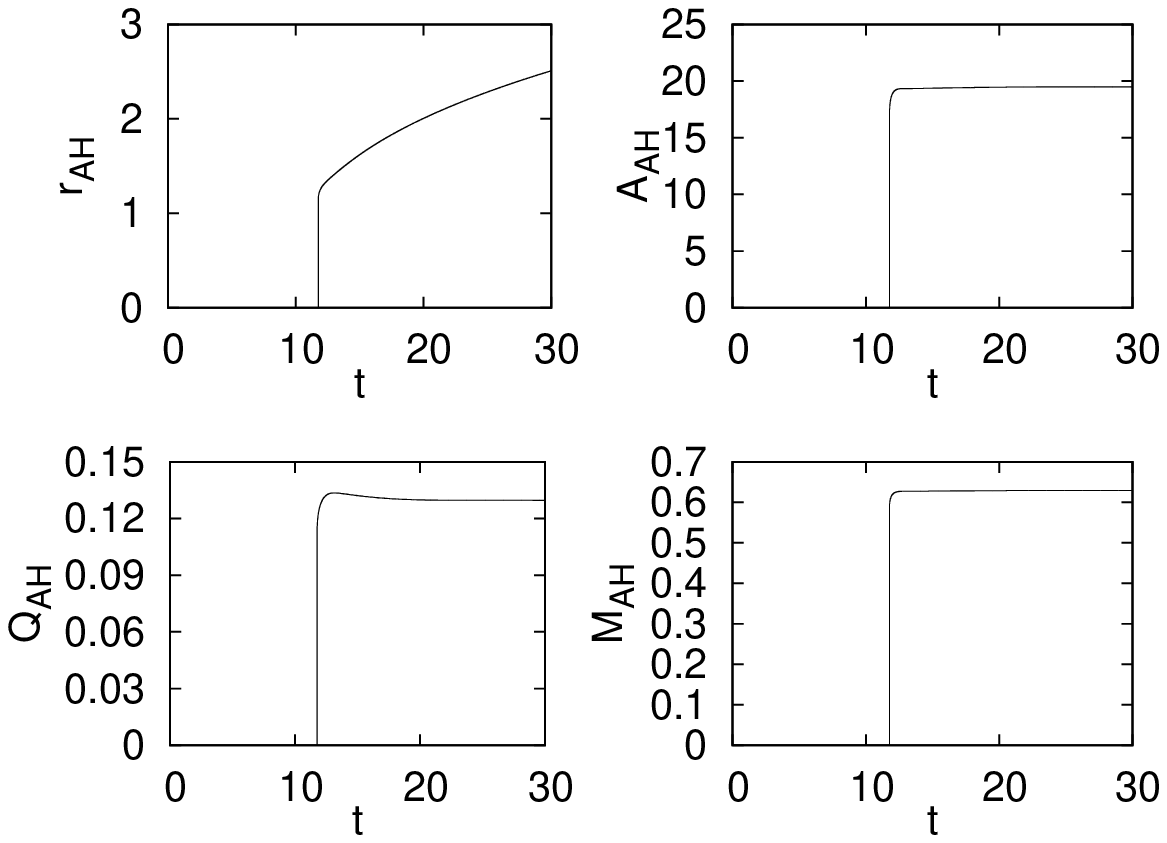}
\caption{Evolution of the apparent horizon for a configuration with
  zero initial charge density and \mbox{$\vphi_0=0.05$},
  \mbox{$q=2.0$}: Coordinate radius (top left), area (top right),
  enclosed charge (bottom left) and horizon mass (bottom right).}
  \label{fig:NQAH}
\end{figure}

For higher values of the amplitude $\vphi_0$ we find that the outcome
is very different from the one described above. To show this we follow
the case with \mbox{$\vphi_0=0.05$} and \mbox{$q=2.0$}
(Figures~\ref{fig:NQEarlyCol}-\ref{fig:NQAH}). The early stage of the
simulation proceeds in the same way as before
(Figure~\ref{fig:NQEarlyCol}), but this time when the incoming pulse
arrives at the origin the lapse function collapses dramatically
(Figure~\ref{fig:NQsnapshotsAlphaCol}), and an apparent horizon is
eventually found at $t \sim 11$ (Figure~\ref{fig:NQAH}), indicating
the formation of a black hole.

Figures~\ref{fig:NQMassCol} and~\ref{fig:NQChargeCol} show the
evolution of the integrated mass and charge, and we see that both tend
to stabilize quickly outside of the apparent horizon after the
dispersed field leaves the central region.  Figure~\ref{fig:NQAH}
shows the evolution of different properties of the apparent horizon.
Since this simulation is done with a vanishing shift vector we see
that the coordinate radius of the apparent horizon keeps growing as
the simulation goes on. This growth of the coordinate radius where the
apparent horizon is located is a well known gauge effect coming from
the fact that the Eulerian (normal) observers are falling and there is
no shift vector to compensate for this.  On the other hand, the area,
enclosed charge, and mass associated with the horizon rapidly
stabilize.  These quantities are well defined on the apparent horizon:
the charge of the black hole is calculated with
equation~\eqref{eq:charge3} evaluated on the apparent horizon radius,
while the total mass is calculated with equation~\eqref{eq:bhmass2}
which includes the contribution due to the enclosed charge.

\begin{figure}[htbp]
\centering
\includegraphics[width=9cm]{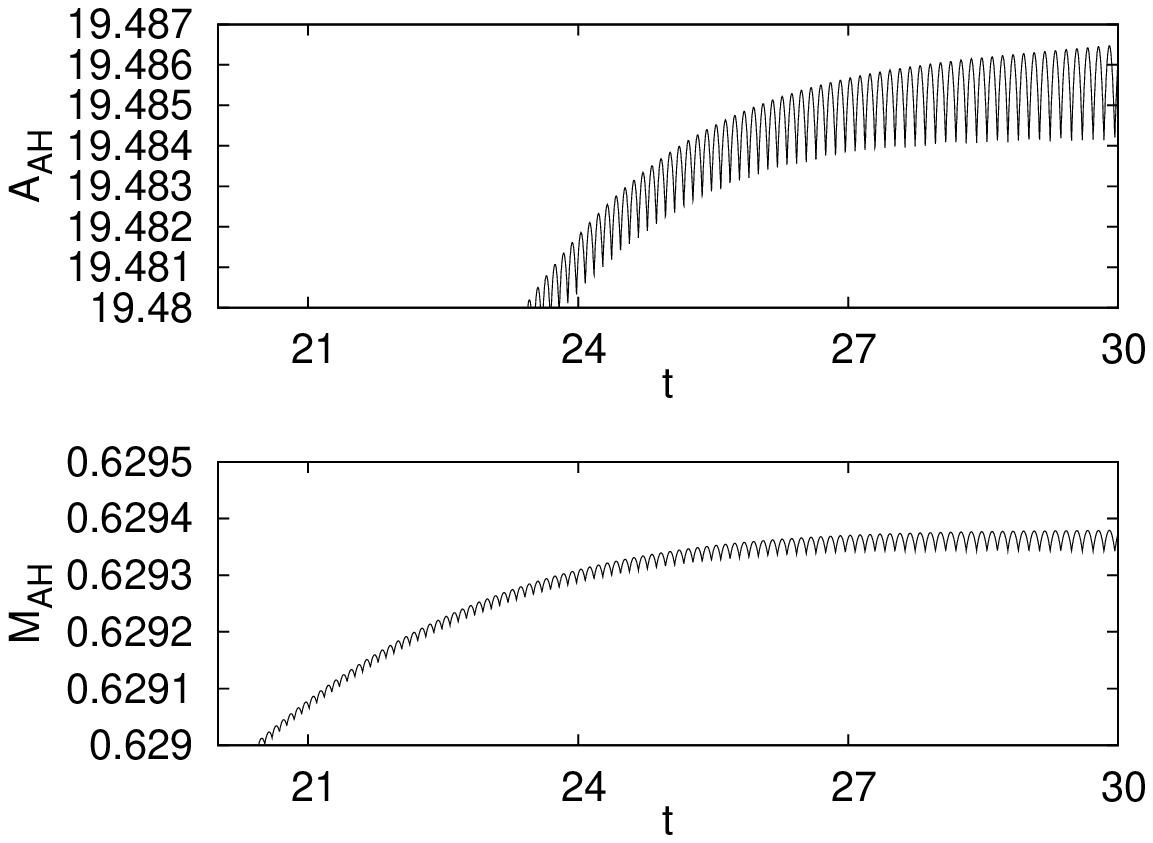}
\caption{Close up of the apparent horizon area (top), and mass
  (bottom) shown in Figure~\ref{fig:NQAH} above. The small
  oscillations are due to numerical error and decrease at higher
  resolutions.}
  \label{fig:NQAH2}
\end{figure}

Ultimately, the lack of a shift vector leads to slice
stretching effects on the metric components that eventually cause our
numerical simulations to fail. However, we find that these effects are
delayed as the mass of the final black hole increases and the
simulations last long enough for us to study the physical properties of
the final black hole.  Figure~\ref{fig:NQAH2} is a close up of the
area and mass of the apparent horizon that shows how these properties
stabilize before the simulation fails (the small oscillations are due
to the numerical method and decrease at higher resolutions).

Once the scalar field outside of the apparent horizon is radiated away
we end up with an electro-vacuum region where only the Coulomb field of
the trapped scalar field remains. The gauge conditions, and in
particular the vanishing shift vector, prevent us from reaching a
stationary situation outside of the trapped surface, but as we have
seen the relevant physical properties rapidly approach stationary
values.

\vspace{5mm}

\begin{figure}[htbp]
\centering
\includegraphics[width=9cm]{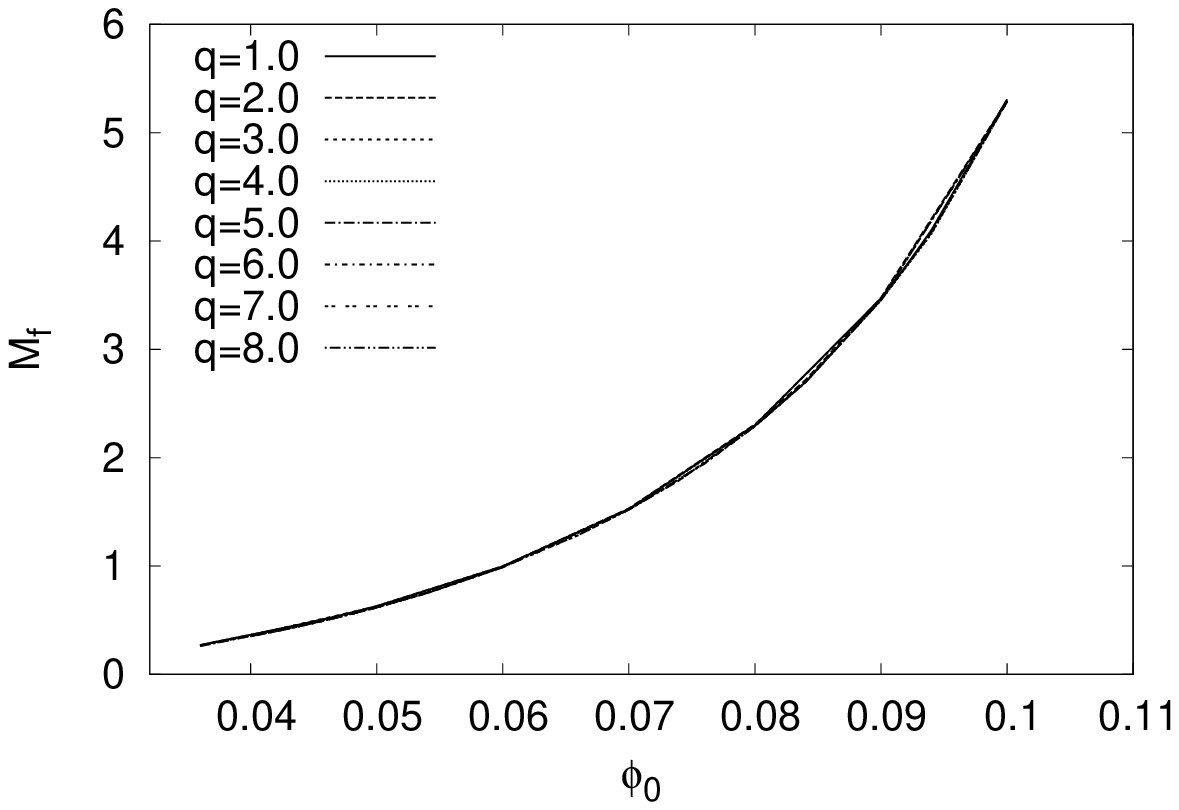}
\caption{Final mass of the black hole $M_f$ for the configurations
  with zero initial charge density, as a function of the initial
  amplitude of the pulse $\vphi_0$ for different values of $q$. Notice
  that for small values of $\vphi_0$ no black hole forms.}
\label{fig:uMf}
\end{figure}

\begin{figure}[htbp]
\centering
\includegraphics[width=9cm]{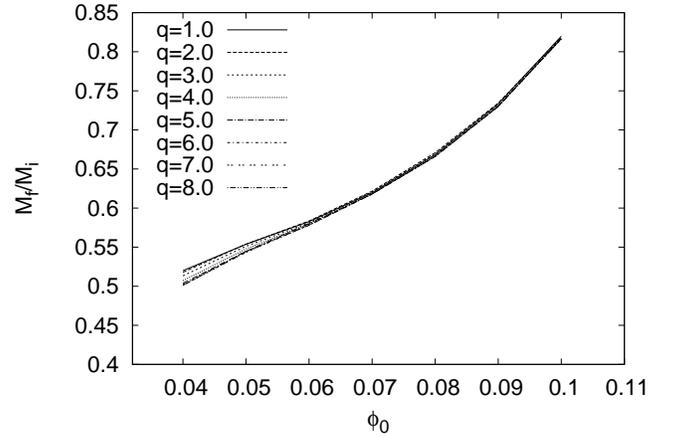}
\caption{Ratio of the final mass of the black hole $M_f$ to the
  initial ADM mass $M_i$ for the configurations with zero initial
  charge density, as a function of the initial amplitude
  $\vphi_0$.}
\label{fig:uMfMi}
\end{figure}

We have also performed simulations for values of the scalar field
charge $q$ ranging from 0.0 to 8.0, and initial amplitude $\vphi_0$
from 0.03 to 0.1, focusing on the region of the parameter space that
leads to configurations that undergo gravitational collapse. The
simulations proceed in a similar way to the one analyzed before, the
main difference is that when increasing both parameters the shape of
the initial pulses gets much more distorted after they split, showing
little resemblance to the initial superposed pulses. In each case,
once an apparent horizon is found its physical properties stabilize
quickly. Also, after the remaining scalar field is radiated away, the
integrated mass and charge stabilize in the electro-vacuum region.  It
is interesting to note that that in this region of the parameter space
the final mass of the black hole is not very sensitive to the value of
the fundamental charge $q$ , which is somewhat surprising since this
mass includes contributions from the electric field (see
Figure~\ref{fig:uMf}).  We have also found that the ratio of final
mass of the black hole to the initial ADM mass of the configuration,
$M_f/M_i$, increases with the amplitude $\vphi_0$ (see
Figure~\ref{fig:uMfMi}), which is easily understood since by
increasing the initial amplitude we get a more compact configuration.

\begin{figure}[htbp]
\centering
\includegraphics[width=9cm]{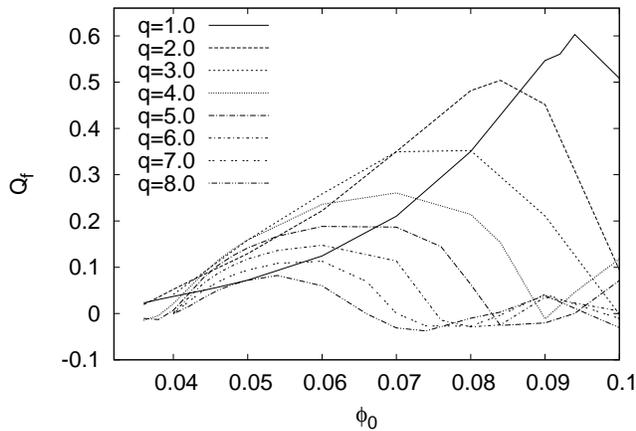}
\caption{Final charge of the black hole $Q_f$ for the configurations
  with zero initial charge density, as a function of the initial
  amplitude $\vphi_0$ for different values of $q$.}
\label{fig:uQf}
\end{figure}

We now turn to the analysis of the charge of the final configurations,
which is somewhat less intuitive. For fixed values of the fundamental
charge $q$ the final charge of the configuration increases initially
with the amplitude of the initial pulse, but it eventually reaches a
maximum and decreases again, and might even oscillate around zero (see
Figure~\ref{fig:uQf}). We find that as we increase the fundamental
charge this maximum is in fact reached for lower initial amplitudes
and has lower a value. 

\begin{figure}[htbp]
\centering
\includegraphics[width=9cm]{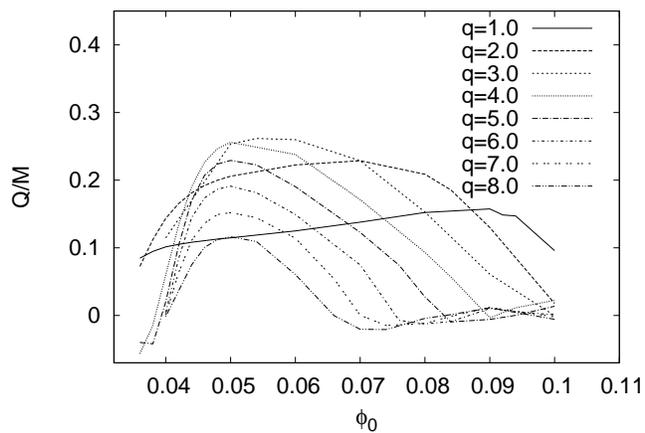}
\caption{Charge to mass ratio $Q_f/M_f$ of the final black hole for
  the configurations with zero initial charge density, as a function
  of the initial amplitude $\vphi_0$ for different values of
  $q$.}
\label{fig:uQM}
\end{figure}

By combining these results we can calculate the quotient of the final
charge and mass of the black hole $Q/M$ (see Figure~\ref{fig:uQM}).
We can observe that as we increase the initial amplitude $\vphi_0$ for
fixed $q$, this ratio reaches a maximum and then decreases again.  The
global maximum that we have found for all our different simulations
corresponds to the value $Q/M \simeq 0.25$, which shows that the final
state of all these configurations is very far from approaching an
extreme charged black hole with $Q/M=1$. 

\subsection{Globally charged configurations}
\label{sec:resultscharged}

For the configurations with non-vanishing global charge we use the
following initial profiles for the scalar field and the scalar
potential
\begin{eqnarray}
\label{eq:initphi}
{\rm Re}{(\vphi)} &=& \vphi_0 \left[ e^{-(r-r_0)^2/\sigma^2}
+ e^{-(r+r_0)^2/\sigma^2} \right] \: ,\\
{\rm Im}{(\vphi)} &=& 0 \: , \\
\Phi&= &\Phi_0 \left[ e^{-(r-r_0)^2/\sigma^2}
+ e^{-(r+r_0)^2/\sigma^2} \right] \: .
\end{eqnarray}

For all the simulations discussed below the pulses are centered
$r_0=5.0$ with a width $\sigma=1.0$. We have also found that by taking
the initial amplitude of the electric potential to be $\Phi_0=1$, it
was possible to construct configurations that have initially a charge
to mass ratio greater than one.

\begin{figure}[htbp]
  \centering
  \includegraphics[width=9cm]{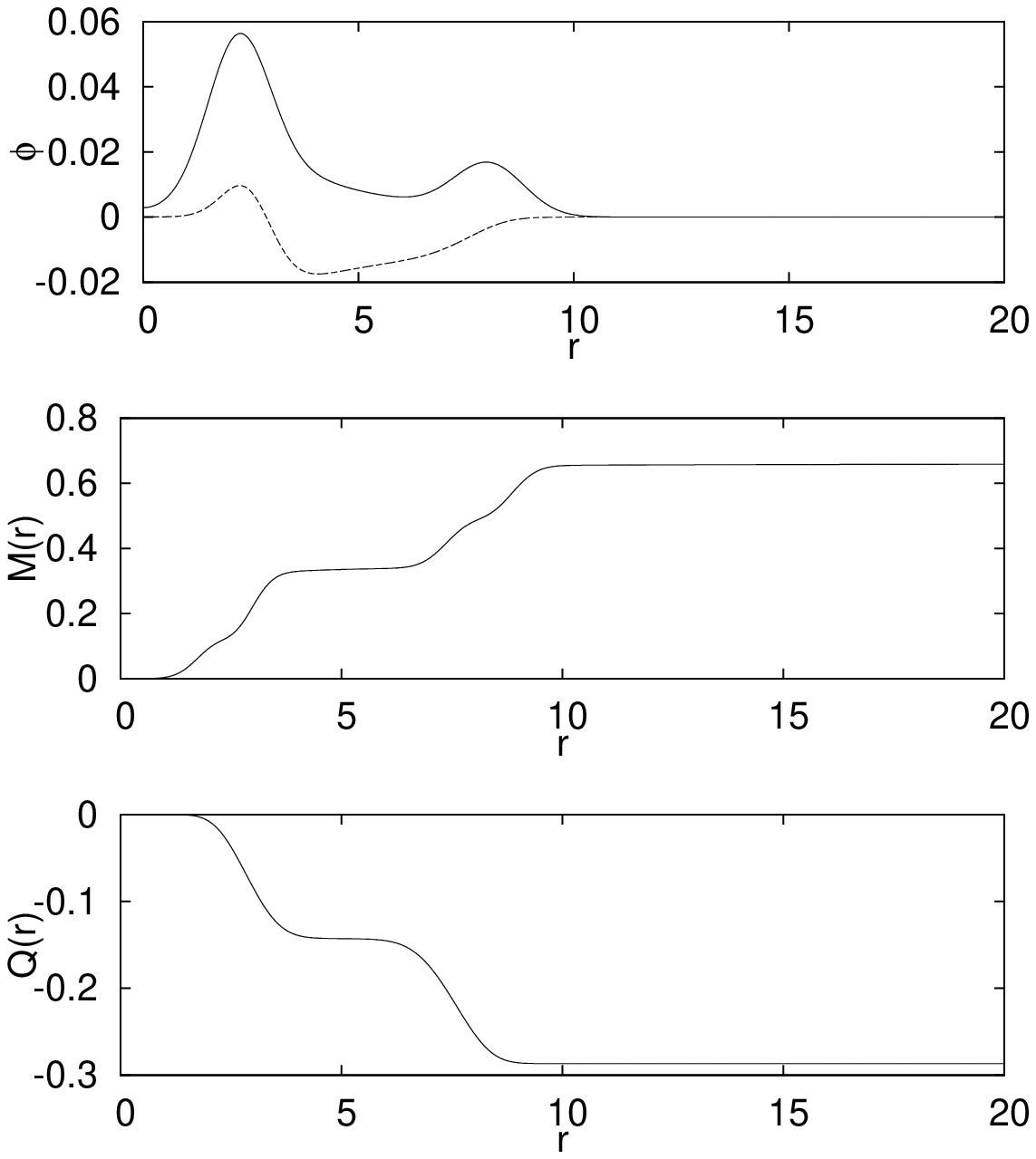}
  \caption{Plot of the early stage of the evolution ($t=4.0$)
    of a configuration with non-vanishing initial charge density and
    $\vphi_0=0.05$, $q=0.5$. The top panel shows the
  scalar field (with the solid and dotted lines corresponding to the
  real and the imaginary parts), while the lower two panels show the
  integrated mass $M(r)$ and charge $Q(r)$.}
  \label{fig:Qq_5_a_05_early}
\end{figure}

\begin{figure}[htbp]
  \centering
  \includegraphics[width=9cm]{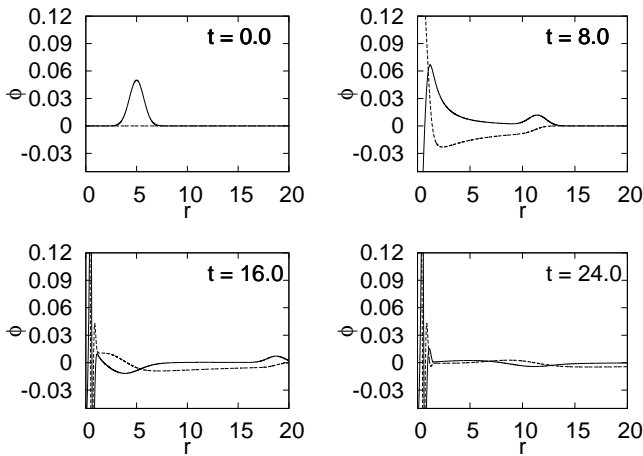}
  \caption{Evolution of the real (solid line) and imaginary (dotted
    line) parts of the scalar field at different times, of a
    configuration with non-vanishing initial charge density and
    $\vphi_0=0.05$, $q=0.5$.}
\label{fig:QsnapshotsScalarCol}
\end{figure}

\begin{figure}[htbp]
  \centering
  \includegraphics[width=9cm]{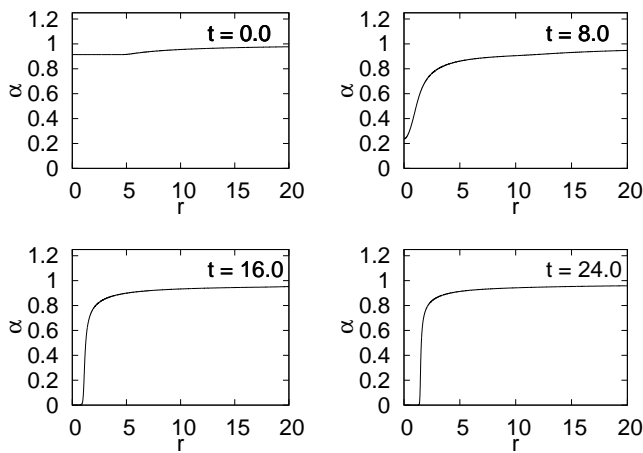}
  \caption{Evolution of the lapse $\alpha$ of a configuration with
    non-vanishing initial charge density and $\vphi_0=0.05$, $q=0.5$.}
\label{fig:QsnapshotsAlphaCol}
\end{figure}

\begin{figure}[htbp]
  \centering
  \includegraphics[width=9cm]{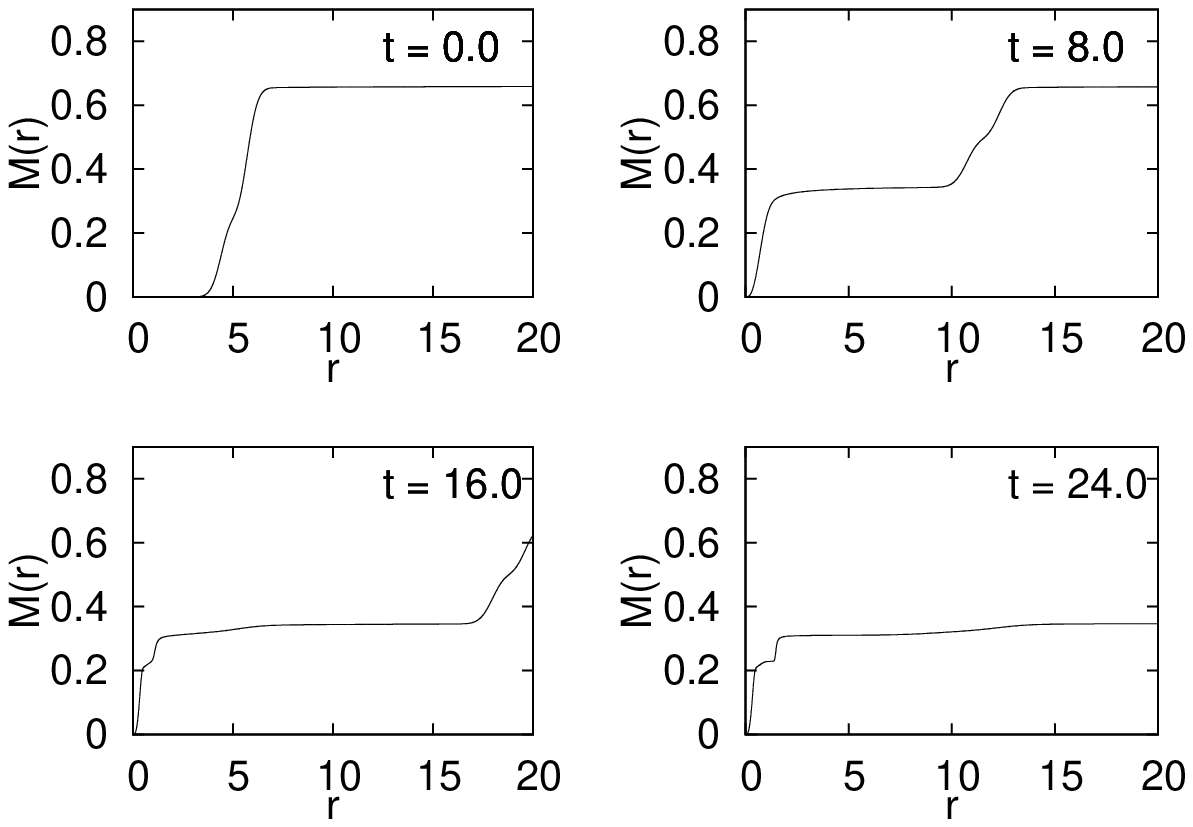}
  \caption{Evolution of the integrated mass $M(r)$ of a
    configuration with non-vanishing initial charge density and
    $\vphi_0=0.05$, $q=0.5$.}
  \label{fig:QMassCol}
\end{figure}

\begin{figure}[htbp]
  \centering
  \includegraphics[width=9cm]{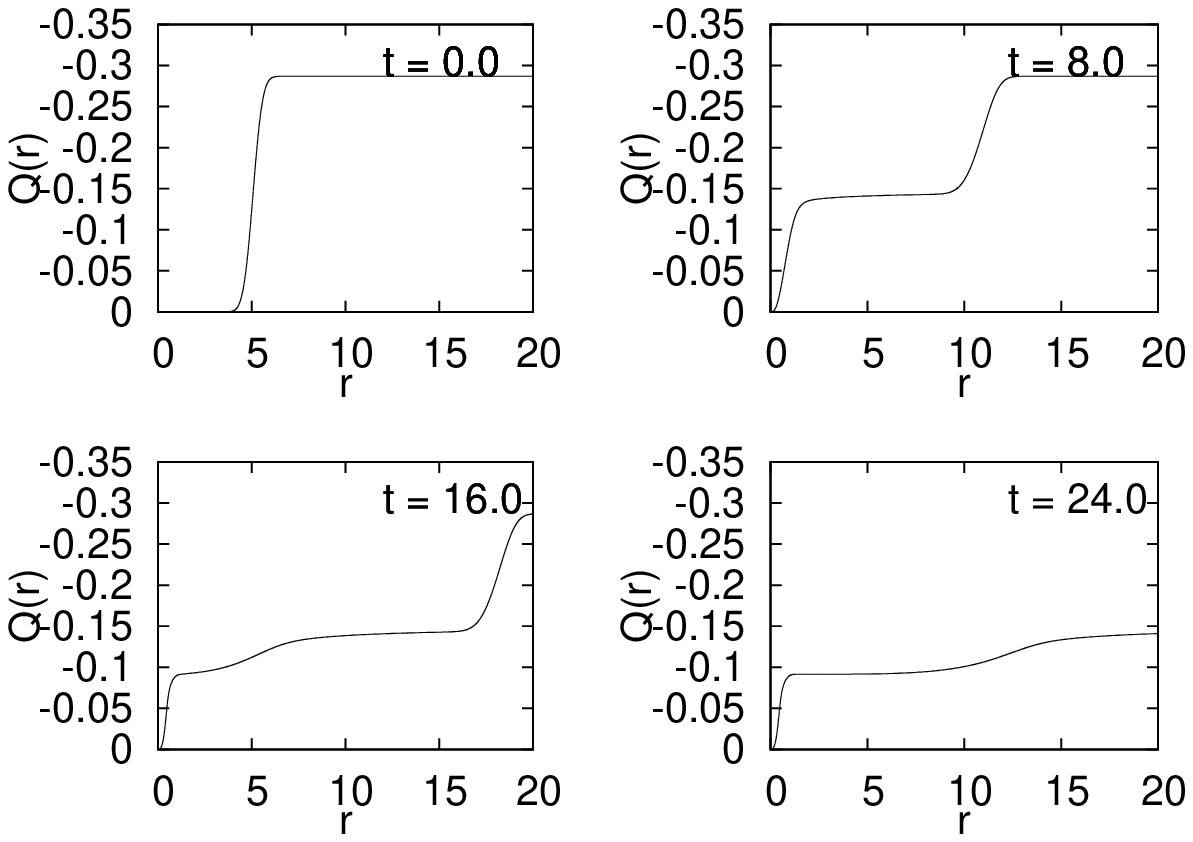}
  \caption{Evolution of the integrated charge $Q(r)$ of a
    configuration with non-vanishing initial charge density and
    $\vphi_0=0.05$, $q=0.5$.}
  \label{fig:QChargeCol}
\end{figure}

\begin{figure}[htbp]
  \centering
  \includegraphics[width=9cm]{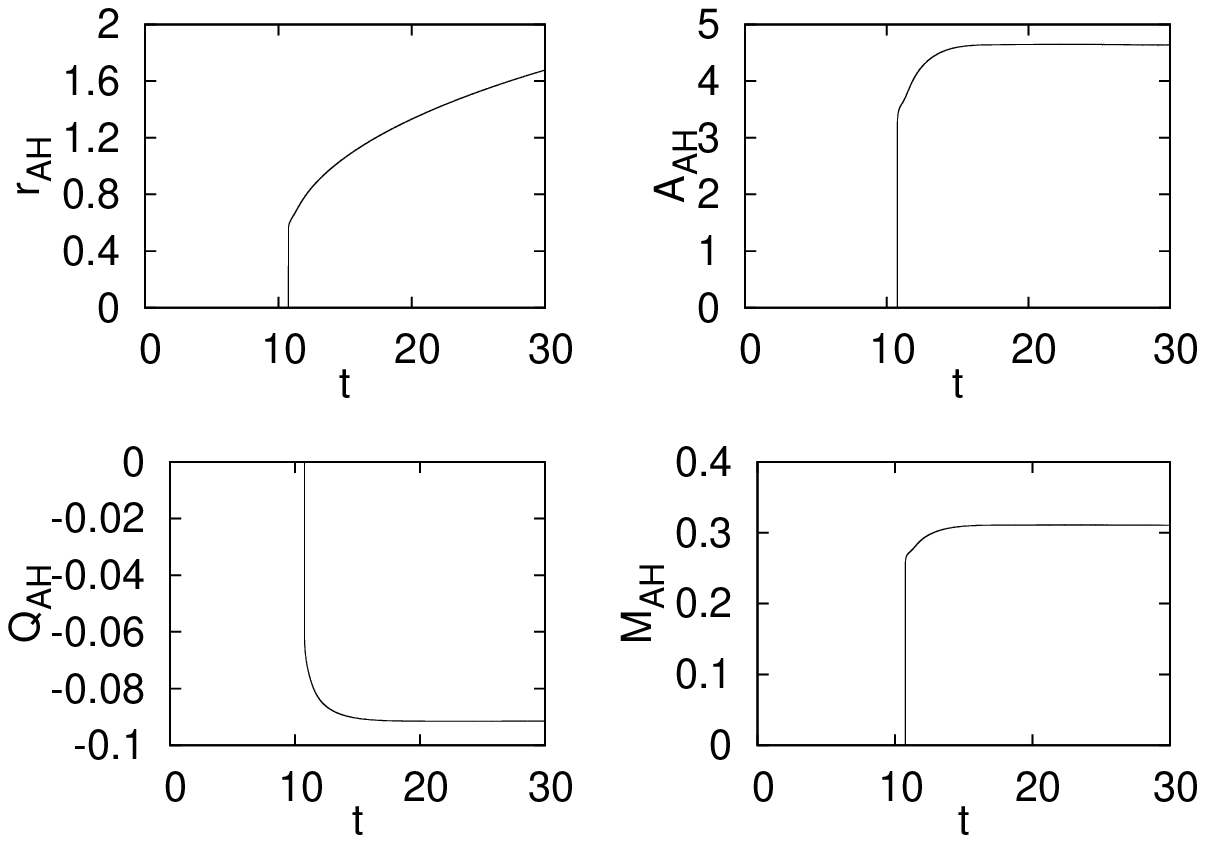}
  \caption{Evolution of the apparent horizon of a configuration
    with non-vanishing initial charge density and $\vphi_0=0.05$,
    $q=0.5$: Coordinate radius (top left), area (top right), enclosed
    charge (bottom left) and horizon mass (bottom right).}
  \label{fig:QAH}
\end{figure}

The evolution of this type of initial data is very similar to the case
with vanishing global charge. As an illustrative example we consider
the case with $\vphi_0=0.05$ and $q=0.5$, which is a combination of
parameters that results in a configuration that undergoes
gravitational collapse.  As can be seen in
Figure~\ref{fig:Qq_5_a_05_early}, which shows the early stages of the
evolution at $t=4$, the initial pulse separates into its
incoming and outgoing components, each one retaining approximately
half of the original charge.
Figures~\ref{fig:QsnapshotsScalarCol}-\ref{fig:QChargeCol} show
snapshots of the scalar field and lapse function, as well as the
integrated mass and charge. Again we see that when the incoming pulse
arrives at the origin the lapse function collapses. Even though this
collapse takes place, some of the scalar field is still dispersed (see
figure~\ref{fig:QsnapshotsScalarCol}) taking away with it part of the
mass and the charge of the configuration, as can be seen in
Figures~\ref{fig:QMassCol} and~\ref{fig:QChargeCol}. Some of the
properties of the apparent horizon for this evolution are shown on
Figure~\ref{fig:QAH}.

\vspace{5mm}

\begin{figure}[htbp]
  \centering
  \includegraphics[width=9cm]{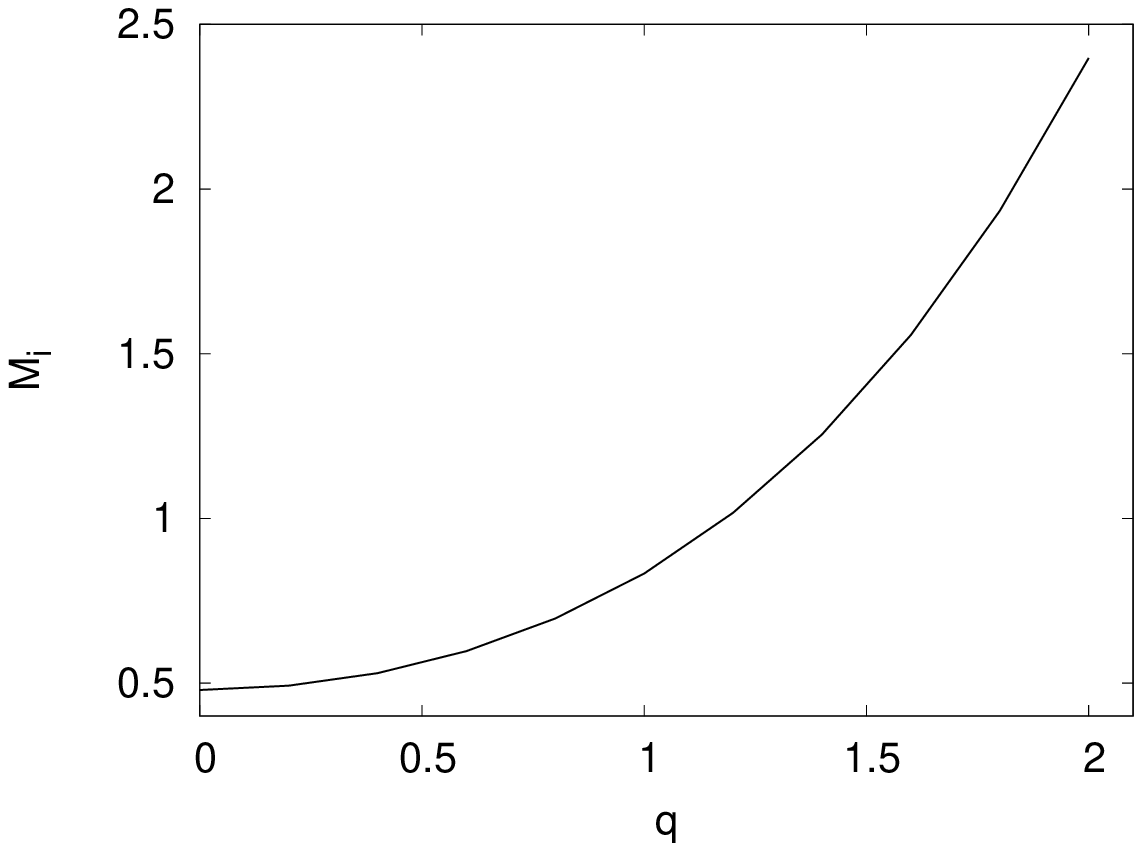}
  \caption{Initial ADM mass for the configurations with non-vanishing
    electric charge, as a function of the fundamental charge $q$ for
    fixed $\vphi_0=0.05$.}
  \label{fig:qMi}
\end{figure}

\begin{figure}[htbp]
  \centering
  \includegraphics[width=9cm]{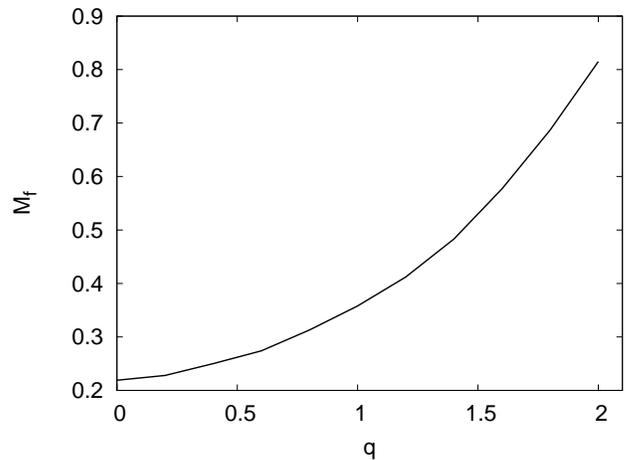}
  \caption{Mass of the final black hole for the configurations with
    non-vanishing electric charge, as a function of the fundamental
    charge $q$ for fixed $\vphi_0=0.05$.}
  \label{fig:qMf}
\end{figure}

\begin{figure}[htp]
  \centering
  \includegraphics[width=9cm]{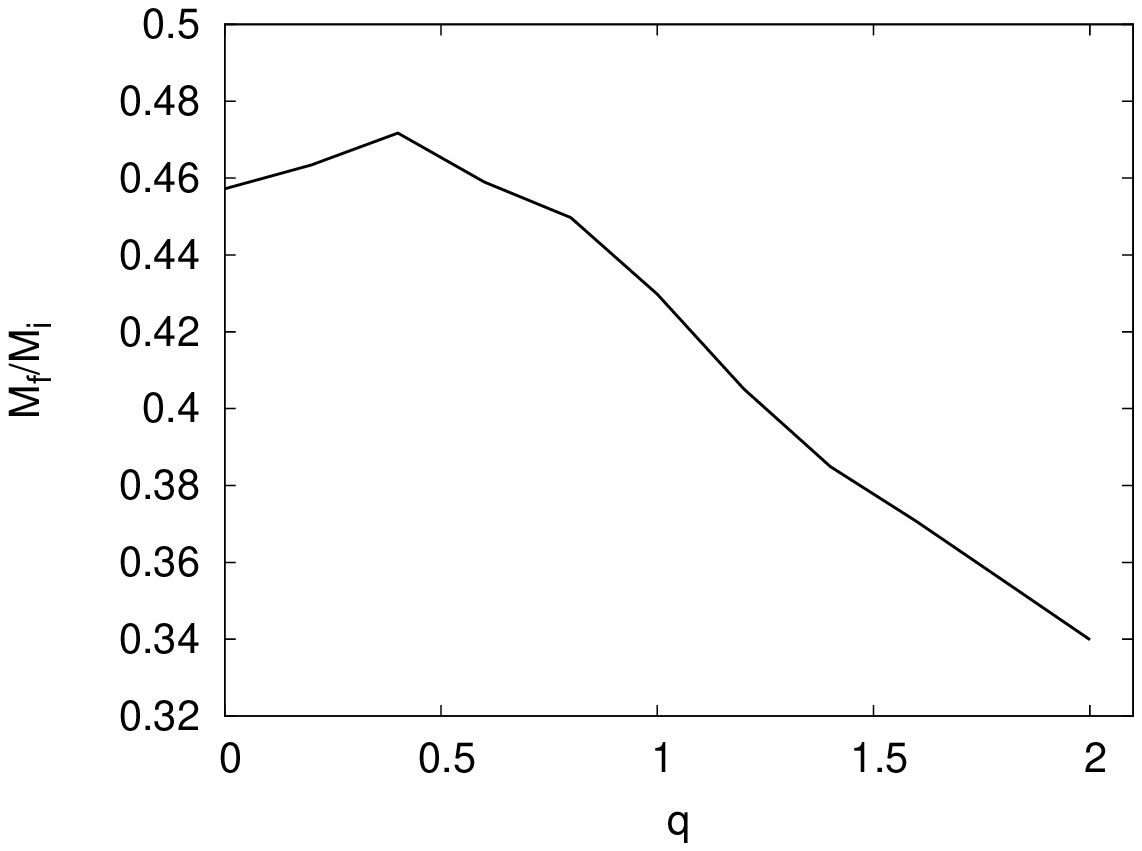}
  \caption{Ratio of the final mass of the black hole to the initial
    ADM mass for the configurations with non-vanishing
    electric charge, as a function of the fundamental
    charge $q$ for fixed $\vphi_0=0.05$.}
    \label{fig:qMfMi}
\end{figure}

We will now concentrate on configurations with amplitude
$\vphi_0=0.05$, and with a fundamental charge that varies from $q=0$
to $q=2$. All these configurations are found to undergo gravitational
collapse. The initial ADM mass of the configurations turns out to be
an increasing function of $q$ as a combined effect of both the
presence of a more intense initial electric field and the larger areal
radius of the initial shell (see Figure~\ref{fig:qMi}). As one would
expect, the mass of the final black hole is also an increasing
function of $q$ (see Figure~\ref{fig:qMf}). Figure~\ref{fig:qMfMi}
shows the ratio of the mass of the final black hole to the initial ADM
mass, $M_f/M_i$. As expected, this is always lower than 0.5 since half
the initial mass is carried away by the outgoing pulse.  Notice also
that the mass ratio shows a strong dependence on the value of $q$, and
in fact decreases for large values of $q$.  This can be interpreted as
an effect of the electric repulsion of the initial pulse: there is an
accumulation of charge of the same sign which repels itself, so that
as $q$ increases a larger fraction of the ingoing pulse in fact ends
up being dispersed to infinity.

\begin{figure}[htbp]
  \centering
  \includegraphics[width=9cm]{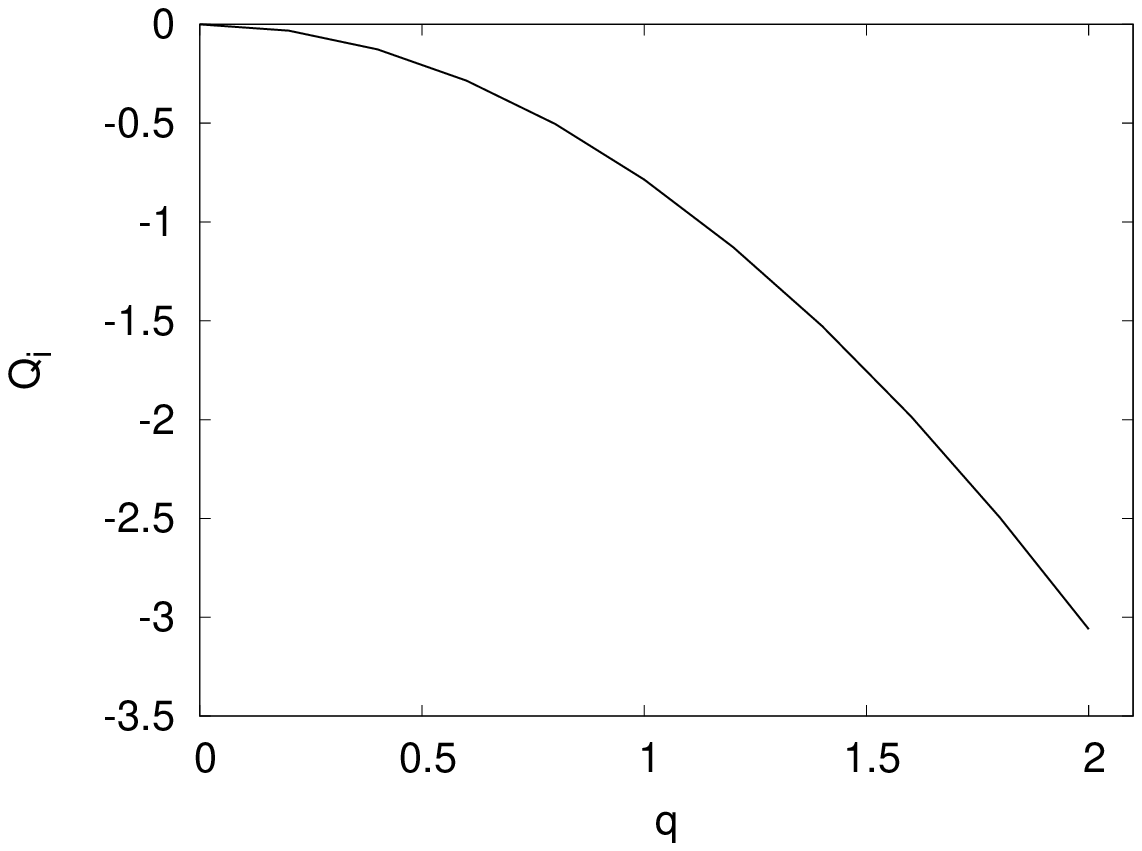}
  \caption{Initial charge of the configurations, as a function of the
    fundamental charge $q$.}
  \label{fig:qQi}
\end{figure}

\begin{figure}[htbp]
  \centering
  \includegraphics[width=9cm]{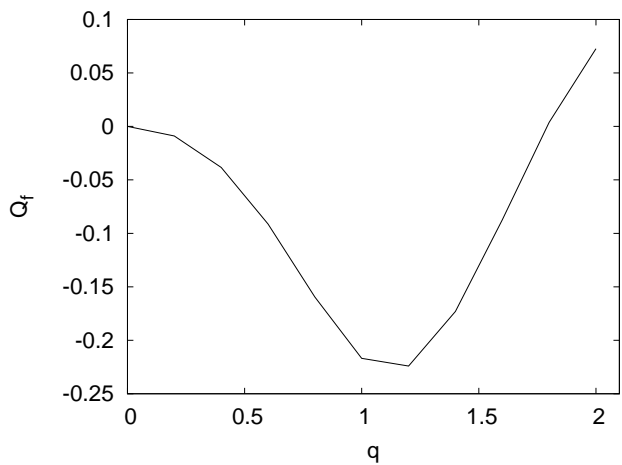}
  \caption{Final charge of the black hole as a function of the
    fundamental charge $q$.}
  \label{fig:qQf}
\end{figure}

\begin{figure}[htbp]
  \centering
  \includegraphics[width=9cm]{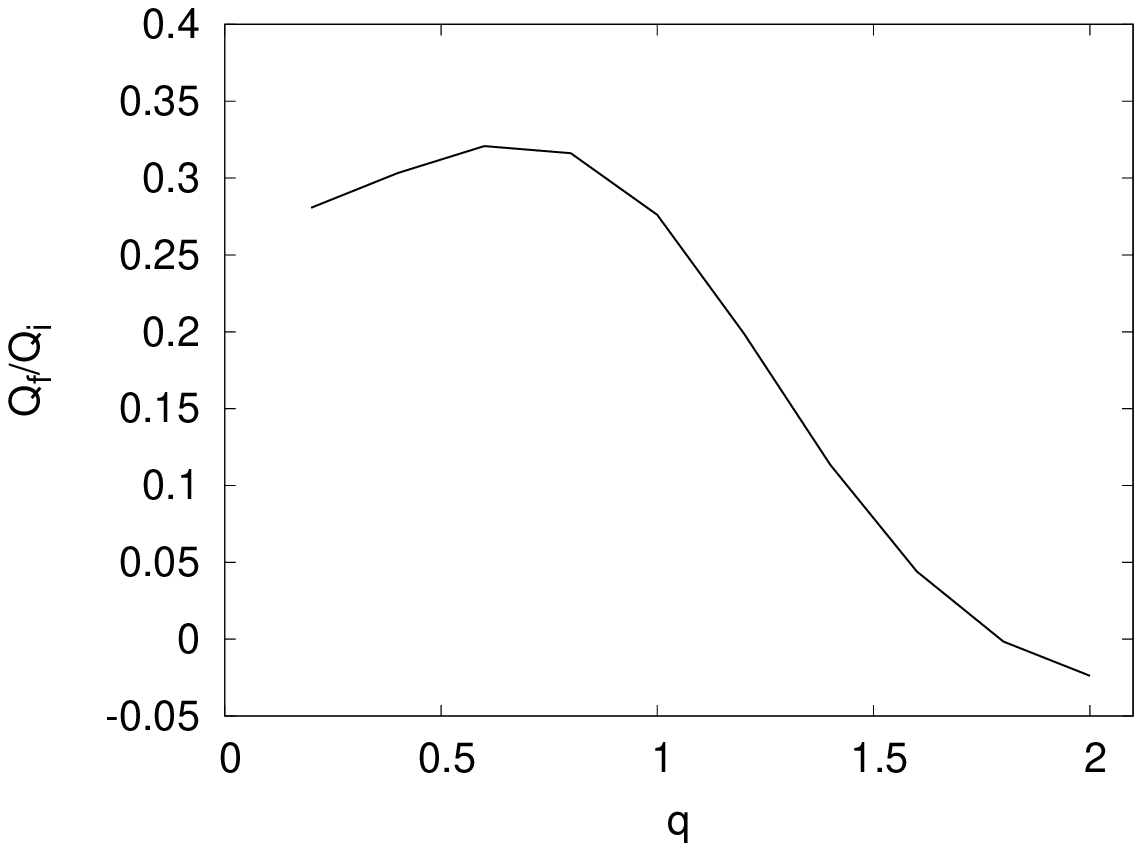}
  \caption{Ratio of the final charge to the initial charge $Q_f/Q_i$,
    as a function of the fundamental charge $q$.}
    \label{fig:qQfQi}
\end{figure}

Considering now the electric charge of these configurations, we find
that the initial charge $Q_i$ depends almost quadratically on the
fundamental charge $q$, as can be seen in Figure~\ref{fig:qQi} (notice
that the initial charge is negative). On the other hand, the final
charge of the black hole $Q_f$ decreases in absolute value for higher
values of $q$ and even changes sign, as shown in
Figure~\ref{fig:qQf}. The ratio $Q_f/Q_i$ is shown in
Figure~\ref{fig:qQfQi}. This behavior can be understood as follows:
when the initial pulse splits the electromagnetic interaction of the
ingoing pulse is so strong that it disperses scalar field that carries
away the excess charge, while it allows accretion of scalar field that
carries electrical charge of the opposite sign, eventually this effect
is so strong that the final charge of the black hole changes sign with
respect to the initial configuration. This change of sign in the
charge of the final black hole is also seen on the initially neutral
configurations (figures \ref{fig:uQf} and \ref{fig:uQM}), but in that
case there is no initial charge to compare with.

\begin{figure}[htbp]
  \centering
  \includegraphics[width=9cm]{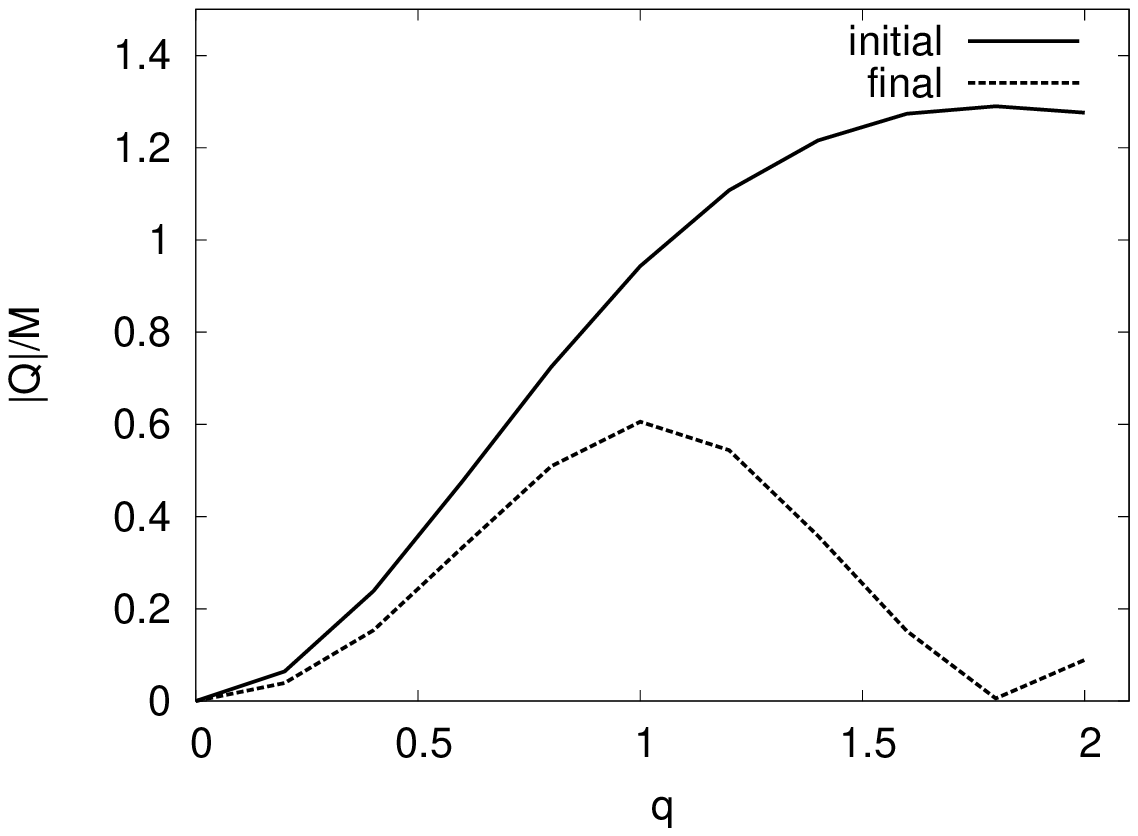}
  \caption{Charge to mass ratio $|Q/M|$ as a function of the
    fundamental charge $q$. The solid line corresponds to the initial
    data, and the dotted line to the final black hole. }
  \label{fig:qQM}
\end{figure}

Finally, in Figure~\ref{fig:qQM} we show the behavior of the charge to
mass ratio $|Q/M|$, both for the initial data and the final black
hole. For the initial data, this ratio grows with the fundamental
charge but eventually flattens at around $q \sim 1.5$. One can notice
that the ratio in fact becomes greater than one beyond $q \sim
1$. However, for the final black hole this ratio is always lower than
its original value, and since the final charge decreases to zero as
the fundamental charge $q$ increases we find that the ratio also
decreases beyond $q=1$. For this family of configurations the maximum
charge to mass ratio for the final black hole is approximately
$|Q_f/M_f| \sim 0.62$ and is attained for a value of the fundamental
charge $q \sim 1$.

\subsection{Nature of the final black hole}
\label{sec:Ressiner}

It is a common feature of the configurations that we have analyzed
that after the collapse of the scalar field in the central region an
apparent horizon forms. We find that even if our gauge conditions
prevent us from approaching a completely stationary situation, the
physical properties of the apparent horizon, essentially its area and
the charge contained within it, settle down quickly. Since we don't
expect the small amount of outgoing scalar field to affect the
collapsed object, we may safely assume that once the apparent horizon
properties settle down the configuration has turned into a black hole,
and that the apparent horizon coincides for all practical purposes
with the event horizon. If this is correct the black hole must be of
the Reissner-N\"ordstrom class, in concordance with the uniqueness
theorems in the case of spherical-symmetry.

One way to verify this is to compare the asymptotic value of the
integrated mass function~\eqref{eq:MassSCH2} with the ``horizon mass''
inferred from the properties of the apparent horizon given
by~\eqref{eq:bhmass2}, $M_{\rm H} = M_{\rm irr} + Q_{\rm H} / 4 M_{\rm
  irr}$.  The numerical error in this case can be estimated from the
amplitude of the small oscillations of these quantities on the higher
resolution runs, as shown on Figure \ref{fig:NQAH2}.  The asymptotic
mass, on the other hand, is measured by reading the value of the mass
function $M(r)$ at a coordinate radius $r=30$ (the midpoint of our
computational domain) at late times, once the final black hole has
settled down. The error in this case can be estimated by considering
the difference of the mass function evaluated at the two highest
resolutions. As we have already mentioned, one should remember the
fact that at late times the region outside the horizon is not in fact
a vacuum since it contains a non-vanishing electric field, so that the
value of the mass function $M(r)$ will always remain smaller that the
ADM mass.  This effect, however, turns out to be very small.

Table~\ref{tab:NQTable} summarizes the results obtained for the family
of initial data with zero initial charge density, for different values
of $q$ and $\varphi_0$.  Notice that, in particular, for the example
shown in the text corresponding to \mbox{$\vphi_0=0.05$},
\mbox{$q=2.0$}, we find an estimated asymptotic mass \mbox{$M_{\infty}
  = 0.6290\pm 0.0003$}, while for the horizon mass we find
\mbox{$M_{\rm H} = 0.6294\pm 0.00007$.}

\begin{table*}[htb]
  \centering
  \begin{tabular}{cc|rrrrrrrr}
    \hline
    \hline
    ~$q$~&~$\vphi_0$~&~$M_{irr}$~&~$Q_{H}$~&~$\delta Q_{H}$~&~$M_{H}$~&~$\delta M_{H}$~&~$M_{\infty}$~&~$\delta M_{\infty}$~&~$Q_{H}/M_{H}$~\\
    \hline
    \hline
    1.0&0.040&0.3656 &0.03698 &0.00003 &0.3666 &0.00008   &0.3651&0.0002  &0.1009       \\
    1.0&0.060&0.9928 &0.1246  &0.00004 &0.9967 &0.00008   &0.9969&0.00005 &0.1250       \\
    1.0&0.080&2.287  &0.3502  &0.0002  &2.300  &0.00003   &2.300 &0.00007 &0.1523       \\
    2.0&0.040&0.3628 &0.0522  &0.00004 &0.3647 &0.00008   &0.3636&0.0003  &0.1431       \\
    2.0&0.050&0.6225 &0.1295  &0.00007 &0.6294 &0.00007   &0.6290&0.0003  &0.2059       \\
    2.0&0.060&0.9856 &0.2221  &0.0002  &0.9982 &0.00007   &0.9983&0.0002  &0.2225       \\
    2.0&0.080&2.282  &0.4821  &0.0003  &2.308  &0.00006   &2.308 &0.00004 &0.2089       \\
    3.0&0.040&0.3596 &0.04136 &0.00004 &0.3606 &0.00008   &0.3604&0.0002  &0.1147       \\
    3.0&0.060&0.9787 &0.262   &0.001  &0.9964 &0.00008   &0.9946&0.0002  &0.2629       \\
    3.0&0.080&2.289  &0.3521  &0.0009  &2.302  &0.00005   &2.303 &0.00007 &0.1530       \\
    4.0&0.040&0.3563 &0.02130 &0.00002 &0.3566 &0.0001    &0.356 &0.0002  &0.05973      \\
    4.0&0.060&0.9769 &0.2363  &0.0008  &0.992  &0.00007   &0.9921&0.0002  &0.2382       \\
    4.0&0.080&2.291  &0.2164  &0.0009  &2.296  &0.00005   &2.297 &0.00009 &0.09425      \\
    5.0&0.040&0.3539 &0.00710 &0.000006&0.354  &0.00008   &0.3537&0.0001  &0.02006      \\
    5.0&0.060&0.9793 &0.1887  &0.0001  &0.9894 &0.00009   &0.989 &0.00008 &0.1907       \\
    5.0&0.080&2.292  &0.06320 &0.0001  &2.293  &0.00005   &2.293 &0.00006 &0.02756      \\
    6.0&0.040&0.3525 &0.0016  &0.000001&0.3525 &0.00008   &0.3525&0.0002  &0.004549     \\
    6.0&0.060&0.9832 &0.1527  &0.002   &0.9897 &0.00006   &0.9895&0.00002 &0.1543       \\
    6.0&0.080&2.292  &-0.0287 &0.00001 &2.292  &0.00005   &2.292 &0.00003 &-0.01252     \\
    7.0&0.040&0.3519 &0.0004  &0.000001&0.3518 &0.00008   &0.3519&0.0001  &0.001137     \\
    7.0&0.060&0.9873 &0.1134  &0.0001  &0.991  &0.00007   &0.991 &0.00003 &0.1144       \\
    7.0&0.080&2.292  &-0.0276 &0.0002  &2.292  &0.00005   &2.292 &0.00005 &-0.01204     \\
    8.0&0.040&0.3519 &-0.0002 &0.00009 &0.3519 &0.00007   &0.3519&0.0003  &-0.0005683   \\
    8.0&0.060&0.992  &0.0603  &0.0003  &0.9928 &0.00007   &0.993 &0.00003 &0.06074      \\
    8.0&0.080&2.291  &-0.0101 &0.00007 &2.291  &0.00005   &2.292 &0.00004 &-0.004409    \\
    \hline
    \hline
  \end{tabular}
  \caption{Properties of the final black hole ($M_{\rm irr}, Q_{\rm
      H}, M_{\rm H}$), and estimated asymptotic mass $M_\infty$, for
    the initial data family with zero initial charge density, for
    different values of $q$ and $\varphi_0$. }
  \label{tab:NQTable}
\end{table*}

The results for the family of initial data with non-vanishing charge
density are also shown in Table~\ref{tab:QTable}, for different values
of $q$ and a fixed value of \mbox{$\varphi_0=0.05$}.  For the example
discussed in the text corresponding to \mbox{$q=0.5$}, we find an
estimated asymptotic mass of \mbox{$M_\infty = 0.3107\pm 0.000576$}
and an horizon mass \mbox{$M_{\rm H} = 0.3109\pm 0.000092$}.

\begin{table*}[htb]
  \centering
  \begin{tabular}{c|rrrrrrrr}
    \hline
    \hline
    ~$q$~&~$M_{irr}$~&~$Q_{H}$~&~$\delta Q_{H}$~&~$M_{H}$~&~$\delta M_{H}$~&~$M_{\infty}$~&~$\delta M_{\infty}$~&~$Q_{H}/M_{H}$~\\
    \hline
    \hline
    % 0.0&0.2463 &0.0     &0.2463&0.000096&0.2465&0.000212&0.0              \\
    % 0.1&0.24909&-0.00293&0.2491&0.00010 &0.2493&0.000284&0.0117623        \\
    % 0.2&0.25735&-0.01223&0.2575&0.000101&0.2578&0.000289&0.0474951        \\
    % 0.3&0.27011&-0.02919&0.2709&0.000096&0.2711&0.000403&0.107752         \\
    % 0.4&0.28573&-0.05526&0.2884&0.000094&0.2888&0.000374&0.191609         \\
    % 0.5&0.30401&-0.09151&0.3109&0.000092&0.3107&0.000576&0.294339         \\
    % 0.6&0.3229 &-0.1378 &0.3376&0.000085&0.3368&0.000833&0.408175         \\
    % 0.7&0.34335&-0.1913 &0.3700&0.000075&0.3686&0.000946&0.517027         \\
    % 0.8&0.37001&-0.2448 &0.4105&0.000063&0.4076&0.000753&0.596346         \\
    % 0.9&0.41722&-0.2788 &0.4638&0.000085&0.4629&0.000405&0.601121         \\
    % 1.0&0.51005&-0.2697 &0.5457&0.000092&0.5450&0.00023 &0.494228         \\
    % 1.1&0.67872&-0.1913 &0.6922&0.000095&0.6939&0.000846&0.276365         \\
    0.0&0.2463 &0.0     &0.0     &0.2463&0.0001  &0.2465&0.0002  &0.0              \\
    0.1&0.2491 &-0.00293&0.000007&0.2491&0.0001  &0.2493&0.0003  &0.01176          \\
    0.2&0.2574 &-0.0122 &0.00003 &0.2575&0.0001  &0.2578&0.0003  &0.04750          \\
    0.3&0.2701 &-0.0292 &0.00006 &0.2709&0.0001  &0.2711&0.0004  &0.1078           \\
    0.4&0.2857 &-0.0552 &0.00009 &0.2884&0.00009 &0.2888&0.0004  &0.1916           \\
    0.5&0.3040 &-0.0915 &0.0001  &0.3109&0.00009 &0.3107&0.0006  &0.2943           \\
    0.6&0.3229 &-0.1378 &0.0002  &0.3376&0.00009 &0.3368&0.0008  &0.4082           \\
    0.7&0.3434 &-0.1913 &0.0002  &0.3700&0.00008 &0.3686&0.0009  &0.5170           \\
    0.8&0.3700 &-0.2442 &0.0003  &0.4105&0.00006 &0.4076&0.0008  &0.5963           \\
    0.9&0.4172 &-0.2785 &0.0004  &0.4638&0.00009 &0.4629&0.0004  &0.6011           \\
    1.0&0.5101 &-0.2695 &0.0003  &0.5457&0.00009 &0.5450&0.0002  &0.4942           \\
    1.1&0.6787 &-0.1913 &0.0002  &0.6922&0.0001  &0.6939&0.0008  &0.2764           \\
    \hline                            
    \hline                            
  \end{tabular}                       
  \caption{Properties of the final black hole ($M_{\rm irr}, Q_{\rm
      H}, M_{\rm H}$), and estimated asymptotic mass $M_\infty$, for
    the initial data family with non-vanishing charge density, for
    different values of $q$ and  $\varphi_0=0.05$. }
  \label{tab:QTable}                  
\end{table*}                          

\vspace{5mm}

Looking at the tables we see that in all cases there is an excellent a
agreement between the asymptotic mass $M_\infty$ and the horizon mass
$M_{\rm H}$ , with the differences falling within the errors
associated with the discretization scheme employed. This result is
quite satisfactory since it compares the asymptotic behavior of the
metric with local measurements on the horizon (its area) and the
integrated charge, and is therefore a strong indicator that the final
black hole is indeed of the Reissner-N\"ordstrom type.

\section{Conclusions}
\label{sec:conclusions}

We have made a systematic study of the collapse of self-gravitating
spherically symmetric configurations of charged scalar field in the
3+1 formalism. To solve the full Einstein-Maxwell-Klein-Gordon (EMKG)
system we coupled the field equations of both the electromagnetic and
scalar fields to a generalized version of the
Baumgarte-Shapiro-Shibata-Nakamura (BSSN) formulation of general
relativity~\cite{Brown:2009dd,Alcubierre:2010is}.

Our main goal in this study was to explore different collapse
scenarios in order to test the cosmic censorship hypothesis and
analyze the mechanisms that make it hold. We focused on two types of
initial data configurations, both of them conformally flat and
time-symmetric: the first set possessed zero initial charge density,
while the other one possessed zero initial current density. Since the
critical collapse of this kind of configurations is well
understood~\cite{Petryk05}, we focused on configurations that lead to
collapse far from the critical regime.  In all such cases, we found
that an apparent horizon forms, giving empirical evidence for the
validity of cosmic censorship hypothesis.

We also studied the final charge to mass ratio of the collapsing
configurations, in order to study how close one can get to an extremal
charged black hole with $|Q|/M=1$. For the initially uncharged
configurations we performed a series of simulations for different
values of the initial amplitude $\varphi_0$ and fundamental charge
$q$. The maximum charge to mass ratio found for the final black hole
with this type of initial data was of $|Q|/M \sim 0.26$, well below
the extremal value of 1. Since these configurations are insensitive
initially to the value of the fundamental charge $q$, there are some
aspects of the dynamics that are almost unaffected.  In particular,
the interaction between the ingoing and outgoing components of the
field is minimal, so the final mass of the black holes also shows
little dependence on the fundamental charge.

The effect of the electromagnetic interaction is better appreciated
when considering the charge of the black hole: increasing the values
of $\phi_0$ and $q$ leads to a larger electromagnetic repulsion which
tends to neutralize the final charge of the collapsed object.  For the
configurations with non-zero global charge the results are very
similar.  The maximum charge to mass ratio for the final black hole
was found to be $|Q|/M \sim 0.6$.  The main difference with the former
family of simulations is that in this case the final mass of the black
hole depends heavily on the fundamental charge $q$. The presence of an
initial electric field produces a considerable electric repulsion of
the initial shell that tends to disperse it.  However, for
intermediate values of $q$ in the region we have explored, this
repulsion is not strong enough to have a significant neutralizing
effect by the time an apparent horizon forms, leading to black holes
with a larger charge to mass ratio than those of the initially
uncharged set.

Our results are consistent with the fact that when considering charged
matter the cosmic censorship conjecture holds generically. In our
case, a strong electromagnetic interaction gives rise to a
redistribution of the fields that avoids the concentration of electric
charge in small regions. This leads to configurations of collapsed
matter that are nearly neutral, and the effect turns out to be greater
as we increase the electromagnetic interaction by going to higher
values of the fundamental charge $q$.  This seem to indicate that
overcharged configurations will not be able to collapse while
retaining the charge excess.

We also focused on studying the properties of the final stationary
black holes. The final configurations observed are consistent with the
uniqueness theorems of spherically symmetrical spacetimes in the sense
that, once the collapse takes place and the charge excess is radiated
away along with the dispersed scalar field, they reduce to the
Reissner-N\"ordstrom geometry outside the horizon. This was verified
by comparing the mass function obtained from the metric expressed in
Boyer-Lindquist coordinates, and the black hole mass associated to the
horizon properties.  Of course, the question as to what happens inside
the horizon is also very interesting.  Eternal Reissner-N\"ordstrom
black holes have a very complex causal structure inside the horizon,
with Cauchy horizons, timelike singularities and wormholes connecting
an infinite repetition of exterior regions.  How exactly this
structure changes when instead of an eternal black hole one considers
a collapse scenario is something that one would like to address
through numerical simulations such as those presented here.  However,
approaching the singularity in a numerical simulation is far from
trivial as dynamical quantities diverge there, and the gauge
conditions used for our present study are not adequate for the task
(we are using a singularity avoiding slicing condition with no shift).

As a final comment, in this work we only considered the case of a
charged massless scalar field, the question of whether the same
results will hold in support when considering a massive scalar field
(or a more general self-interaction potential) is still open. It is
well known that the dynamics of the scalar field changes drastically
when considering a non-trivial potential, and that gravitationally
bound configurations can be found in that case, which could presumably
lead to higher charge to mass ratios for the collapsing scenarios.

~

\begin{acknowledgments}

The authors wish to thank Dario N\'u\~nez and Marcelo Salgado for many
useful discussions and comments.  This work was supported in part by
Direcci\'on General de Estudios de Posgrado (DGEP-UNAM), by CONACyT
through grant 82787, and by DGAPA-UNAM through grants IN113907 and
IN115310. J.M.T. also acknowledges a CONACyT postgraduate scholarship.

\end{acknowledgments}

\bibliographystyle{bibtex/prsty}
\bibliography{bibtex/referencias}

\end{document}